\documentclass[aps,twocolumn,prb,showpacs,preprintnumbers,superscriptaddress,amsmath,amssymb,floatfix]
{revtex4-1}
\usepackage[charter]{mathdesign}	
\usepackage{amsmath}	
\usepackage{graphicx}	
\usepackage{dcolumn}	
\usepackage{bm}		
\usepackage[english]{babel}
\usepackage{verbatim}
\usepackage{color}
\usepackage[percent]{overpic}
\newcommand{\upmu}{\relax{\mbox{\usefont{U}{psy}{m}{n}{m}}}}
\newcommand{\uplambda}{\relax{\mbox{\usefont{U}{psy}{m}{n}{l}}}}%

\usepackage[normalem]{ulem}
\newcommand{\se}[1]{\textcolor{cyan}{#1}}

\begin{document}
\preprint{APS/123-QED}

\author{S. Esser}
\affiliation{Experimentalphysik VI, Center for Electronic Correlations and Magnetism,
Augsburg University, D-86159 Augsburg, Germany}
\author{C. F. Chang}
\affiliation{Max Planck Institut f\"ur Chemische Physik fester Stoffe, D-01187 Dresden, Germany}
\author{C.-Y. Kuo}
\affiliation{Max Planck Institut f\"ur Chemische Physik fester Stoffe, D-01187 Dresden, Germany}
\author{S. Merten}
\affiliation{I. Physikalisches Institut,
Georg-August-Universit\"{a}t G\"ottingen, D-37077 G\"ottingen, Germany}
\author{V. Roddatis}
\affiliation{Institute of Material Physics,
Georg-August-Universit\"{a}t G\"ottingen, D-37077 G\"ottingen, Germany}
\author{T. D. Ha}
\affiliation{Max Planck Institut f\"ur Chemische Physik fester Stoffe, D-01187 Dresden, Germany}
\author{A. Jesche}
\affiliation{Experimentalphysik VI, Center for Electronic Correlations and Magnetism,
Augsburg University, D-86159 Augsburg, Germany}
\author{V. Moshnyaga}
\affiliation{I. Physikalisches Institut,
Georg-August-Universit\"{a}t G\"ottingen, D-37077 G\"ottingen, Germany}
\author{H.-J. Lin}
\affiliation{National Synchrotron Radiation Research Center, 101 Hsin-Ann Road, Hsinchu 30076, Taiwan}
\author{A. Tanaka}
\affiliation{Department of Quantum Matter, ADSM, Hiroshima University, Higashi-Hiroshima 739-8530, Japan}
\author{C. T. Chen}
\affiliation{National Synchrotron Radiation Research Center, 101 Hsin-Ann Road, Hsinchu 30076, Taiwan}
\author{L. H. Tjeng}
\affiliation{Max Planck Institut f\"ur Chemische Physik fester Stoffe, D-01187 Dresden, Germany}
\author{P. Gegenwart}
\affiliation{Experimentalphysik VI, Center for Electronic Correlations and Magnetism,
Augsburg University, D-86159 Augsburg, Germany}
\email{philipp.gegenwart@physik.uni-augsburg.de}

\title{Strain induced changes of electronic properties of B--site \\ordered double perovskite Sr$_2$CoIrO$_6$ thin films}

\date{
\today%
}

\begin{abstract}
B-site ordered thin films of double perovskite Sr$_2$CoIrO$_6$ were epitaxially grown by a metal-organic aerosol deposition technique on various substrates, actuating different strain states. X-ray diffraction, transmission electron microscopy and polarized far-field Raman spectroscopy confirm the strained epitaxial growth on all used substrates. Polarization dependent Co $L_{2,3}$ X-ray absorption spectroscopy reveals a change of the magnetic easy axis of the antiferromagnetically ordered (high-spin) Co$^{3+}$ sublattice within the strain series. By reversing the applied strain direction from tensile to compressive, the easy axis changes abruptly from in-plane to out-of-plane orientation. The low-temperature magnetoresistance changes its sign respectively and is described by a combination of weak anti-localization and anisotropic magnetoresistance effects.
\end{abstract}


\maketitle
\section{Introduction}

\begin{figure}[t]
\includegraphics[width=0.95\linewidth]{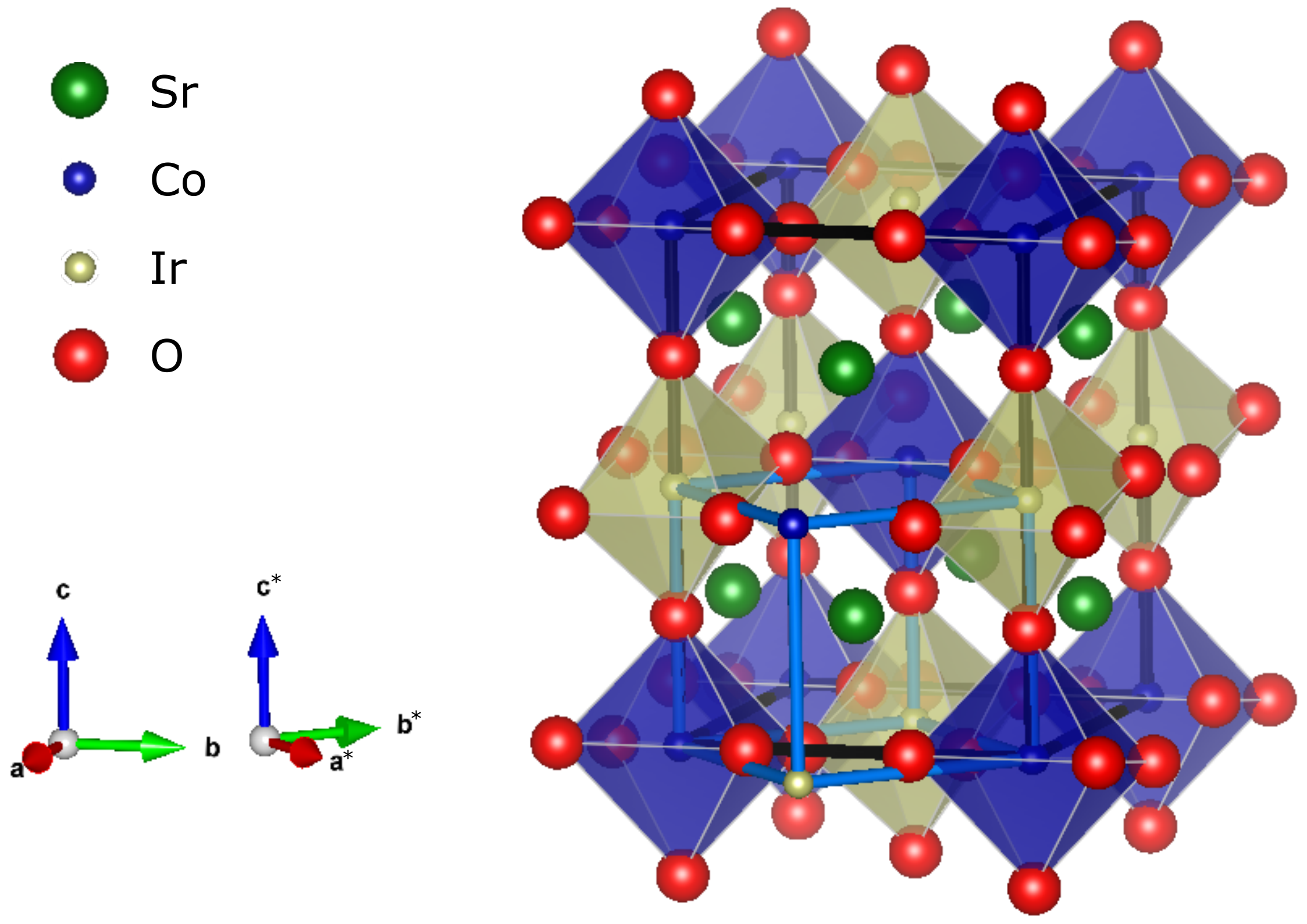}
\caption{\label{Fig_Crystal}Crystal structure of Sr$_2$CoIrO$_6$ visualized with VESTA \cite{VESTA} based on Ref. \cite{Mikhailova}. The light blue cell indicates the pseudo-cubic (pc) unit with rock-salt type ordering at the B-sites. Note the superstructure with alternating Ir and Co layers along the [111]$_\text{pc}$ direction.}
\end{figure}

Double perovskites (DPs) A$_2$BB'O$_6$ display interesting electronic and magnetic properties, strongly depending on the degree of the B-site ordering, which as well is determined both by size and valence mismatch of the involved B-site cations \cite{RevDP,Ohtomo,Anderson}. In addition, epitaxial stabilization in thin films could lead to an improvement of B-site ordering as observed for example in the La$_2$FeCrO$_6$ system \cite{Chakraverty}. The most prominent ordering type of the B-site cations is the rock salt like structure with alternating B and B' planes along the pseudo cubic [111]$_\text{pc}$ direction \cite{RevDP}, shown in Figure \ref{Fig_Crystal} for the title compound of this study. Besides the ferrimagnetic halfmetal Sr$_2$FeMoO$_6$ (Ref. \cite{Sr2FeMoO6}) there are other highly insulating ferromagnetic ordered DPs, like multiferroic La$_2$CoMnO$_6$ (Ref. \cite{La2CoMnO6}) and Ba$_2$CuOsO$_6$ (Ref. \cite{Ba2CuOsO6}), a magnetic insulator recently synthesized und
 er high-pressure.

Recently, correlated oxides with strong spin orbit (SO) coupling attracted great attention. In particular, the SO coupling in iridates with Ir$^{4+}$ ($5d^5$ configuration) ions in octahedral coordination results in four occupied $J_\text{eff} = 3/2$ and two half-filled $J_\text{eff} = 1/2$ states~\cite{Kim}. As was first verified for the  layered perovskite Sr$_2$IrO$_4$ (Ref. \cite{KimScience}), already a moderate Coulomb repulsion is sufficient to induce a SO Mott insulating state, with magnetic $J_\text{eff} = 1/2$ moments. SO coupling is also important for semimetalic iridates \cite{PRL114}. As realized by Jackeli and Khalliulin \cite{kitaev} novel magnetic exchange interactions, in particular the honeycomb Kitaev exchange~\cite{Kitaev-model}, can arise from $J_\text{eff} = 1/2$ magnetic moments. This initiated strong interest in two- and three-dimensional honeycomb iridates~ \cite{arXiv}.

Iridate DPs offer another novel playground to investigate the electronic and magnetic properties arising from the competition of SO coupling, electronic correlations and structural distortions. This leads, for example, in Sr$_2$CeIrO$_6$ to an insulating state with weak antiferromagnetic (AFM) orbital order~ \cite{Sr2CeIrO6}. Also both La$_2$MgIrO$_6$ and La$_2$ZnIrO$_6$ host SO Mott insulating magnetic states~ \cite{La2XIrO6}. Sr$_2$YIrO$_6$ represents a class of DPs with Ir$^{5+}$ configuration and evidence of novel magnetism has been reported~\cite{PRL112}. However, subsequent work \cite{PRB95-Sr2YIrO6}, also on the isoelectronic Ba$_2$YIrO$_6$ \cite{PRB96-Ba2YIrO6} related these observations to diluted paramagnetic impurities.

Strontium-iridate derived DPs may also be interesting from the perspective of tuning the properties of the semimetallic three-dimensional perovskite SrIrO$_3$ to a topological state. A tight-binding model for this material in Ref.~\cite{Carter} revealed a symmetry protected nodal line made of $J_\text{eff}=1/2$ bands below the Fermi level. The same work proposed that the line node can be lifted enabling strong topological insulator behavior if sublayer reflection symmetry could be broken and SrIrO$_3$-based B-site ordered DPs such as Sr$_2$CoIrO$_6$ (SCIO) were suggested as possible route for realization.

There were very few reports~\cite{Mikhailova,Narayanan} on the bulk SCIO, prepared as polycrystalline samples by solid-state reaction from mixtures of oxides and carbonates (SrCO3). SCIO crystallizes in the monoclinic strcuture with the space group I2/m and pseudocubic lattice constants $a=0.3909$~nm, $b=0.3925$~nm and $c=0.3921$~nm. The Co/Ir ordered structure is stabilzed due to a large difference in cation radii between Co$^{2+}$ (0.0745 nm) and Ir$^{4+}$ (0.0625 nm) ions. The antiferromagnetic (AFM) ordering with $T_N \sim 70$~K (AFM Curie-Weiss temperature $\Theta=-138$~K) and very small spontaneous magnetization, $M\approx 0.005 \mu_B$/f.u. was detected. The electron transport was found to by an insulating-like obeying a Mott variable-range hopping scenario with a gap $\sim 0.05$~eV.

Here we report for the first time the growth, structure and electronic properties of SCIO thin films, prepared by metalorganic aerosol deposition (MAD) technique, which employs an oxygen-rich growth atmosphere and enables to prepare high-quality perovskite thin films\cite{SchneiderPRL,Moshnyaga,Jungbauer}. The Co$^{3+}$ high spin state and, consequently, the Ir$^{5+}$ configuration was obtained, that is different as compared to the SrIrO$_3$ perovskite. Another motivating question was whether an incomplete B-site order with about 13.2\,\% site mixing in previously reported bulk SCIO~\cite{Narayanan} could be improved by the in-plane epitaxy strain in thin films and how the electronic and magnetic properties will be influenced by epitaxial strain. Coherently strained SCIO thin films were epitaxially grown on various perovskite substrates and demonstrate a remarkable strain control of magnetotransport. By changing from tensile to compressive strain a sign reversal of the magnetoresistance due to 
 a change of the magnetic easy axis from in- to out-of-plane configuration was observed.

\section{Experimental}
\label{sec:experimental}
Thin films of SCIO were grown by MAD on (111) oriented SrTiO$_3$ (STO), as well as on various pseudo-cubic (pc) (001) oriented substrates.
The lattice mismatch of the used substrates ranges from -1.51\,\% (GdScO$_3$) over -0.99\,\% (DyScO$_3$), 0.05\,\% (STO) and 0.96\,\% (LSAT) to 1.09\,\% (NdGaO$_3$), thus, covering a broad range in both compressive and tensile direction.
To protect the SCIO film an STO capping layer was grown also by MAD directly after the SCIO film.
Phase purity, crystal structure and strain states were determined at room temperature by x-ray diffraction, using a PHILIPS X'PERT diffractometer, operated with Cu-K$_{\alpha1,2}$ radiation. The thickness of the films was determined from XRR measurements, performed by means of a BRUKER D8 ADVANCE diffractometer, and further simulated with the \textsc{ReMagX} \cite{ReMagX} program.

The far-field Raman spectra were measured at room temperature by using a HORIBA Jobin Yvon LabRAM HR Evolution confocal Raman spectrometer in
the back-scattering geometry.
A neodymium-doped yttrium aluminum garnet laser with a wavelength of $\lambda = 532\,\text{nm}$ (second
harmonic generation; Laser Quantum torus 532 with 100\,mW, limited to 1\,mW to avoid heating effects) is appliedfor excitation; the size of the laser spot was $\sim1\,\upmu\text{m}$. 
For the polarization-dependent Raman measurements, the incident linear polarization of the laser can be tuned between P- and S-polarization by rotating a $\uplambda$/2 wave plate. The scattered light polarization is determined by an analyzer with two switchable configurations (P-polarization and S-polarization) in front of the detector.

The high-resolution scanning transmission electron microscopy (HR-STEM) studies were carried out on in a cross-section geometry using a FEI Titan 80-300 environmental transmission electron microscope operated at 300\,kV
and equipped with a Gatan Quantum ER image filter. The cross-section lamella sample was prepared using a focus ion-beam (FIB) machine (FEI Nova NanoLab 600 DualBeam instrument).

The polarization-dependent X-ray absorption spectroscopy (XAS) measurements were performed at the Dragon beamline of the National Synchrotron Radiation Research Center (NSRRC) in Taiwan. The spectra were recorded at 300 K using the total electron yield method (TEY) from SCIO films which were capped with 2 nm STO. The photon energy resolution at the Co $L_{2,3}$ edges was set at 0.3 eV and the degree of linear polarization was 99 \%. The samples were mounted on a holder which was tilted with respect to the incoming beam, such that the Poynting vector of the light makes an angle of 70$^{\circ}$ with respect to the [001]$_\text{pc}$ surface normal. By rotating the sample around this Poynting vector, the polarization of the electric field can be varied continuously from E $\parallel$ 20$^{\circ}$ off the [001]$_{pc}$ surface normal, i.e. E $\parallel$ c (20$^{\circ})$, to E $\perp$ the [001]$_\text{pc}$ surface normal, i.e. E $\parallel$ ab. A CoO single crystal were measured simultaneously in a separate chamber to serve as en
 ergy reference for Co $L_{2,3}$ edge.

 The temperature and field dependences of electrical resistance were measured by means of a Physical Property Measurement System (PPMS) using a four-probe van der Pauw geometry within external measurement setup due to the high resistance, $\sim$ M$\Omega$, of films at low temperatures.
The STO capped samples were micro-structured for the resistance measurement by an \textit{in-situ} optical lithography without any contact to air between the subsequent steps. For this purpose, a special four step method was developed, described in detail in the supplementary material (SM) \cite{SM}.

The temperature and magnetic field dependent magnetization were measured in a 250\,nm thick SCIO film using a Magnetic Property Measurement System (MPMS) equipped with a 7\,T magnet in a stabilized DC mode. For each measuring point the raw background signal of the substrate was carefully subtracted from the measured raw signal of the sample and the resulting signal was analyzed in the standard way of the MPMS.

\section{Results and Discussion}

\subsection{Structural investigation}

\begin{figure}[t]
\includegraphics[width=0.95\linewidth]{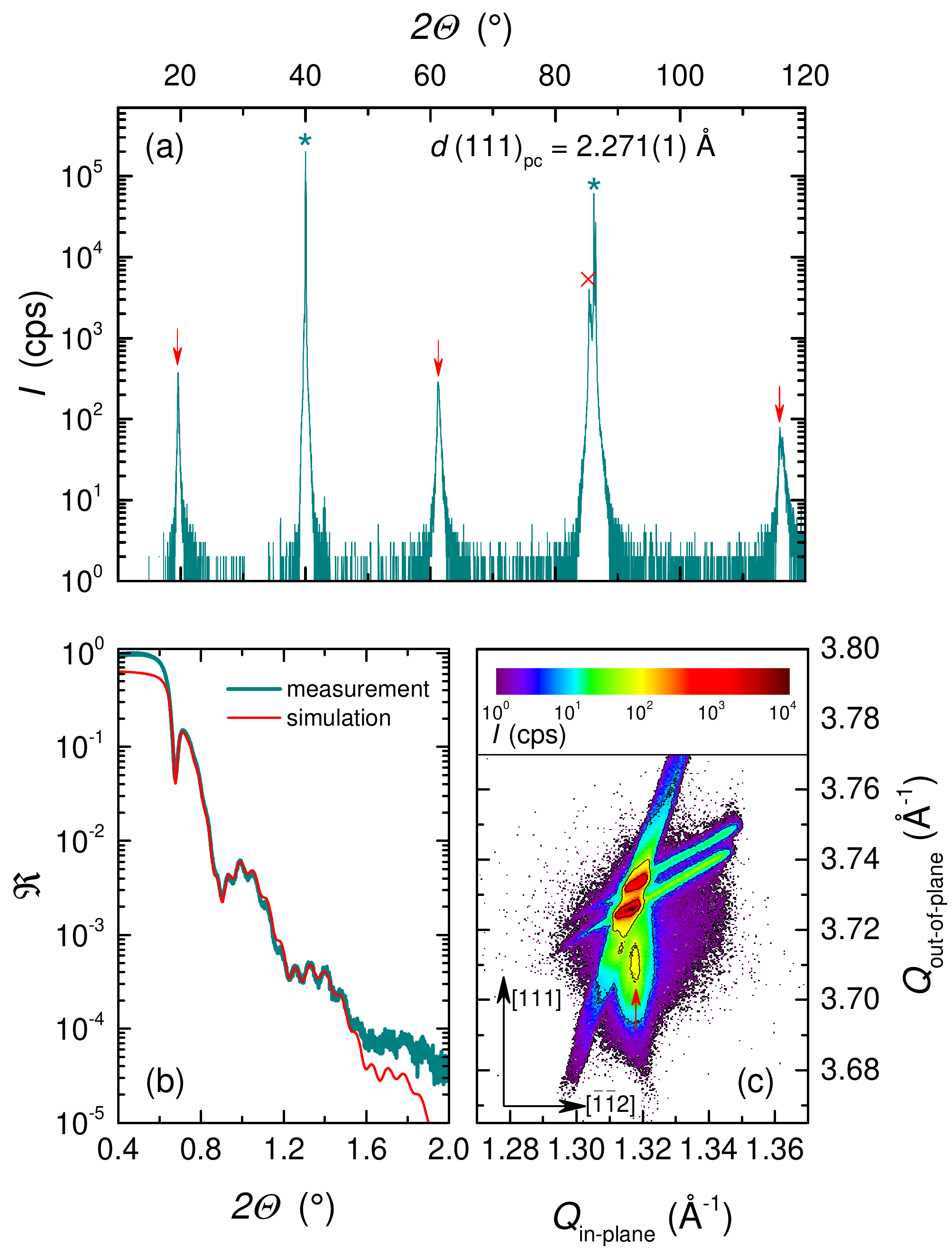}
\caption{\label{Fig_RSM}(Color online) (a) $\theta-2\theta$ XRD scan: Green stars indicating peaks from the (111)-STO substrate, the red cross shows the (222)-SCIO reflection and the arrows mark peaks due to ordered superstructure. (b) XRR measurement of a 84\,nm thick SCIO film with 21\,nm STO capping including simulation with \textsc{ReMagX} \cite{ReMagX}. (c) Reciprocal space map around the (112) peak of STO indicates fully strained thin film (arrow marks SCIO signal). The double peak feature of STO substrate is owed by the Cu-K$_{\alpha 1}$/K$_{\alpha 2}$ doublet.}
\end{figure}

XRD measurements on (111)$_\text{pc}$ oriented thin films on (111)$_\text{c}$ STO (see Figure \ref{Fig_RSM} (a)) indicate an out-of-plane epitaxy. The presence of B-site ordering in films was evidenced by observation of (1/2 1/2 1/2) superstructure reflections, marked by red arrows in Figure \ref{Fig_RSM} (a). The extracted pseudo cubic out-of-plane lattice parameter, $d(111)_\text{pc} = 2.271(1)\,\mathring{\text{A}}$, for all studied SCIO/STO films with $d<250\,\text{nm}$ is slightly larger than that measured for bulk SCIO, $d(111)_\text{pc} = 2.256\,\mathring{\text{A}}$~\cite{Mikhailova}. The reason is the in-plane compressive strain $\sim+0.05\,\%$ due to the lattice mismatch between SCIO and STO. Reciprocal space mapping around the (112)-STO peak (see Figure \ref{Fig_RSM} (c)) confirms the fully strained state of the film. Small angle XRR measurements in Figure \ref{Fig_RSM} (b) indicate a large scale homogeneity of all SCIO films, $d(\text{SCIO}) = 2-250\,\text{nm}$, as 
 well as for the STO capping layer, $d(\text{STO}) = 0 - 20\,\text{nm}$.  The XRR signal can be well simulated with \textsc{ReMagX} \cite{ReMagX} yielding extracted interface roughness of less then 2\,nm.

\begin{figure}[t]
\includegraphics[width=0.95\linewidth]{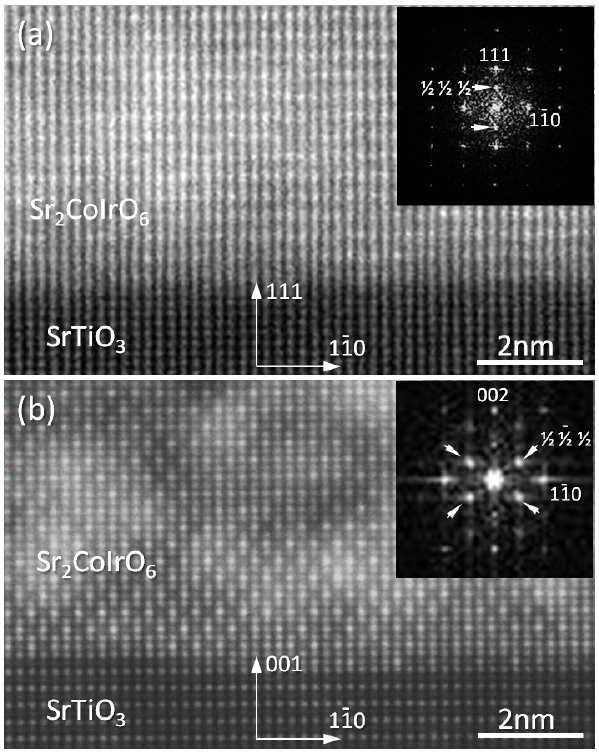}
\caption{\label{Fig_STEM} HAADF HR-STEM images of the interface between SCIO grown on (111) (a) and (001) (b) STO substrate. Both images evidence a fully strained thin film with no defects at the interface. Fourier transformations of the SCIO film (insets) proove the (pseudo) cubic structure with ordered superstructure, indicated by additional $(\frac{1}{2}\frac{1}{2}\frac{1}{2}$)$_\text{pc}$ spots .}
\end{figure}

HR-STEM images and corresponding FFTs in Figure  \ref{Fig_STEM} confirm a high structural quality and  the B-site ordering of SCIO films grown on (111) and (001) STO substrates.  One can clearly see the pseudo-cubic symmetry with extra spots (white arrowheads) along the [111]$_\text{pc}$ direction due to the established B-site ordered superstructure. The SCIO/STO interfaces in look coherent in agreement with results from the reciprocal space map, thus, evidencing a fully strained state of the film. Moreover, no dislocations or other defects are observed. Also the B-site ordering is clearly seen in Figure  \ref{Fig_STEM} (b) since the atomic number of Ir is much larger than the atomic number of Co. The degree of Co/Ir ordering is estimated using HR-STEM image simulation to be more than 65\% (see SM Figure S3) \cite{SM}. 

\begin{figure}[t]
\includegraphics[width=\linewidth]{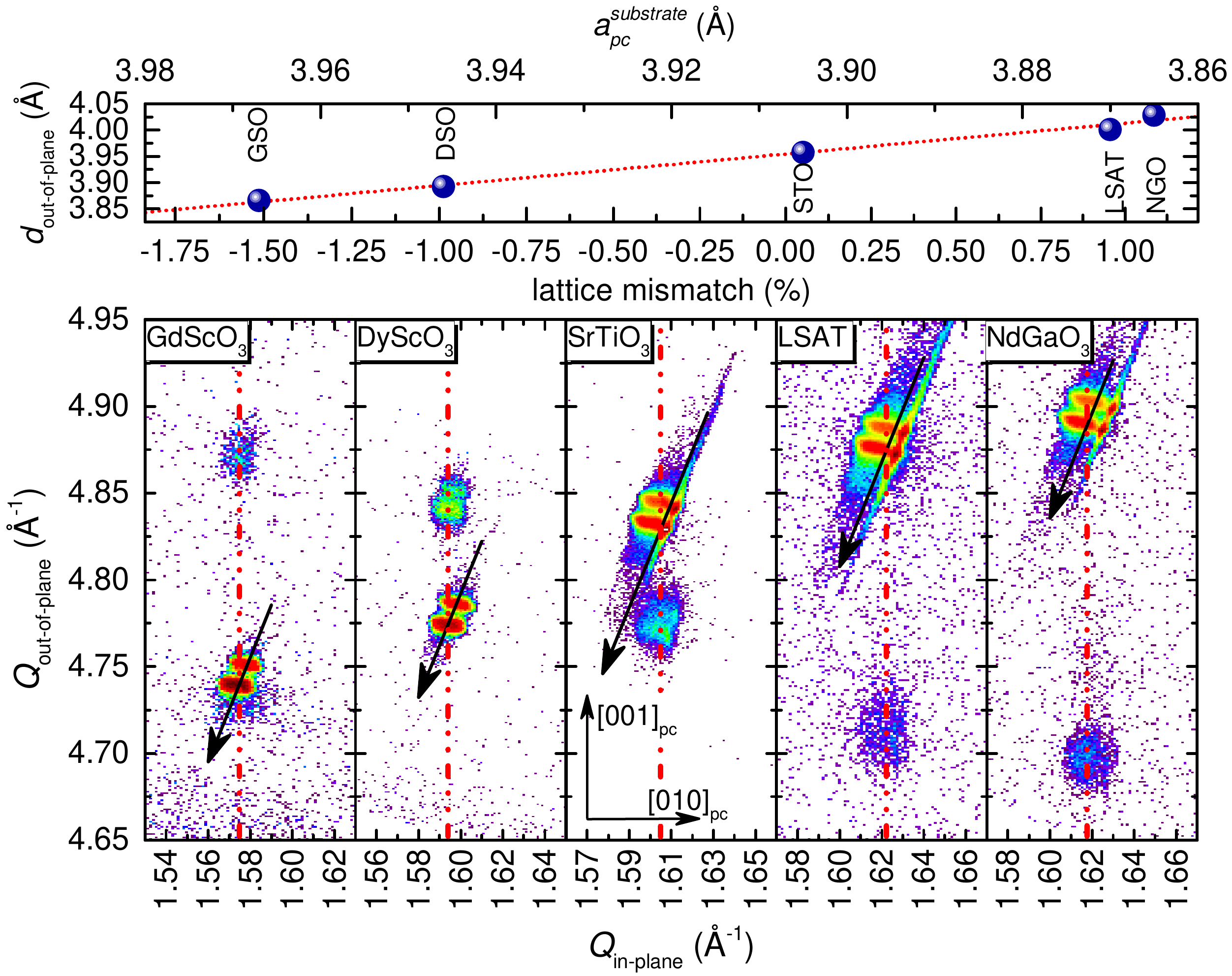}\newline
\includegraphics[width=0.95\linewidth]{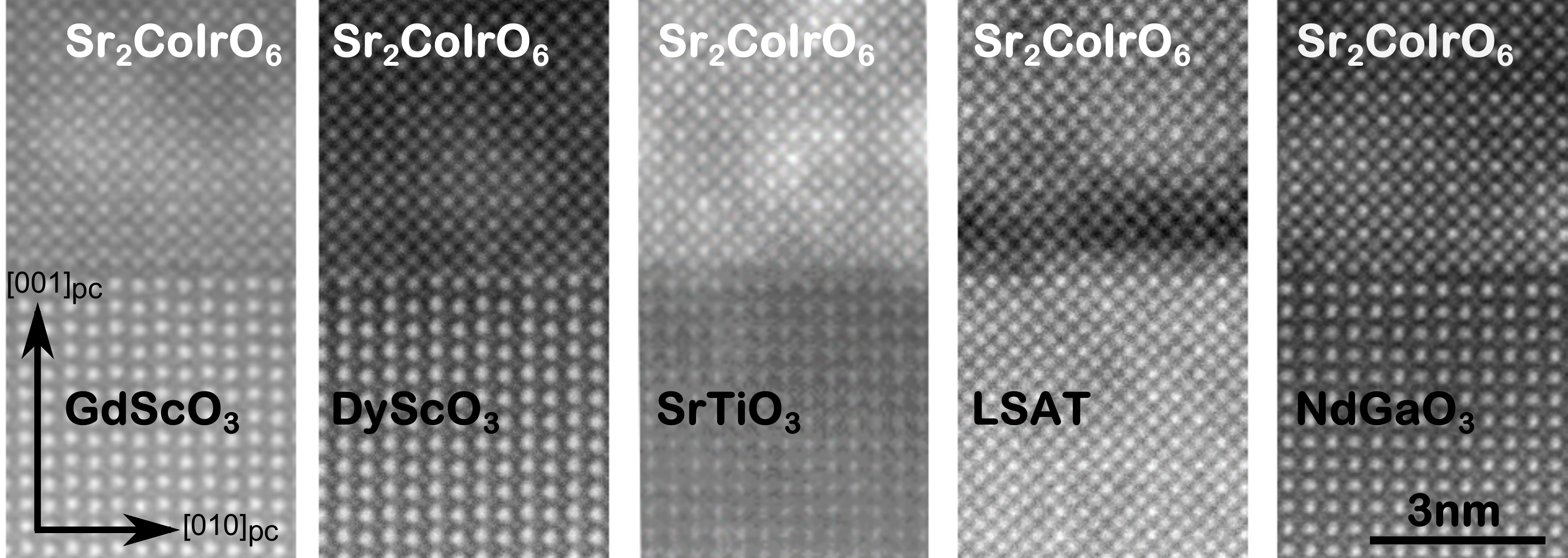}
\caption{\label{Fig_RSMs}(Color online) \textit{Top panel}: Strain dependence of the out-of-plane lattice parameter, red line indicates linear behaviour. \textit{Middle panel}: Reciprocal space maps around the (pseudo) cubic (013)$_\text{pc}$ peak of each used substrate indicating a fully strained thin film in each case. The double peak feature for substrate (and in some cases also for the thin film) reflections is owed by the Cu-K$_{\alpha 1}$/K$_{\alpha 2}$ doublet. The black arrows always point towards the origin. The dashed red lines indicate the adaption of the in-plane lattice parameter due to a constant in-plane scattering vector component. \textit{Bottom panel}: HAADF STEM images of the SCIO--substrate interface for various substrates.}
\end{figure}

To realize a biaxial tensile strain along the [100]$_\text{pc}$ and the [010]$_\text{pc}$ direction, we used substrates of GdScO$_3$ (GSO, $a_\text{pc} = 3.967\,\mathring{\text{A}}$) and DyScO$_3$ (DSO, $a_\text{pc} = 3.946\,\mathring{\text{A}}$) in (001)$_\text{pc}$ orientation. To exert biaxial compressive strain the (001)$_\text{pc}$-oriented SrTiO$_3$ (STO, $a_\text{c} = 3.905\,\mathring{\text{A}}$), LSAT ($a_\text{c} = 3.870\,\mathring{\text{A}}$) and NdGaO$_3$ (NGO, $a_\text{pc} = 3.865\,\mathring{\text{A}}$) were used. 

In the top panel of Figure \ref{Fig_RSMs} the evaluated linear relation between the out-of-plane film lattice parameter $d_\text{out-of-plane}$ and the pseudo-cubic substrate lattice parameter $a^\text{substrate}_\text{pc}$ is shown for STO(20 nm)/SCIO(20 nm) films. Within the linear elasticity theory this behavior indicates a fully strained state of films with the poisson's ratio of $\nu = 0.407(8)$.
Reciprocal space mapping around each (pseudo) cubic (013)$_\text{pc}$ substrate peak (see Figure \ref{Fig_RSMs} (middle panel)) prove the fully strained state of each film grown on the used substrates. In addition, HAADF STEM images also verify the strained film/substrate (001)$_\text{pc}$ interfaces (see Figure \ref{Fig_RSMs} (bottom panel)), preserved up to thickness $\sim50\,\text{nm}$ of SCIO film. 

\begin{figure}[t]
\includegraphics[width=0.925\linewidth]{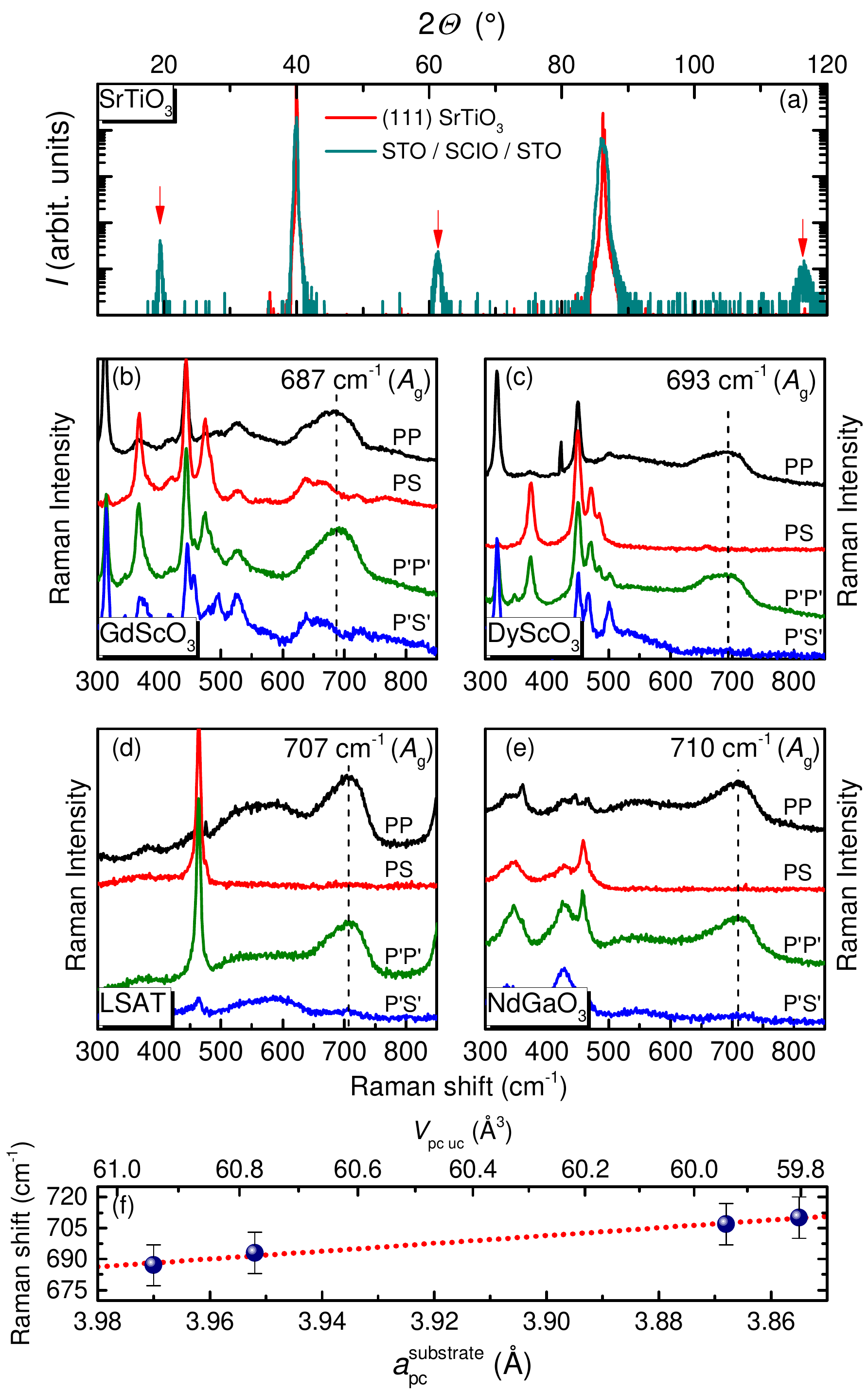}
\caption{\label{Fig_Bsite}(Color online) (a) $\theta-2\theta$ XRD scan in tilted geometry with $\vec{Q}$ parallel to <111>$_\text{c}$ of STO: the arrows mark peaks due to the ordered superstructure. Polarized far-field Raman spectra of SCIO films on (b) GSO, (c) DSO, (d) LSAT and (e) NGO at room temperature. The spectra are taken in the parallel PP (P'P') and the crossed scattering PS (P'S') configurations (with an additional in-plane rotation of $\Phi = 45\,^\circ$ around the [001]$_\text{pc}$ direction). (f) Strain dependence of Raman shift of the breathing mode, line indicates fully strained state with linear dependence.}
\end{figure}

In the case of STO as substrate material the B-site ordering was investigated by a $\theta-2\theta$ XRD scan in tilted geometry with $\vec{Q}$ parallel to <111>$_\text{c}$ of STO and therefore, regarding to the small lattice mismatch between SCIO and STO, also nearly parallel to <111>$_\text{pc}$ of SCIO.
The visible superstructure peaks in the collected XRD pattern (see Figure \ref{Fig_Bsite} (a)) could be well distinguished from the STO background, comparable to the results of the (111)$_\text{pc}$ oriented thin films (see Figure \ref{Fig_RSM} (a)), and indicating well developed B-site ordering.
A larger lattice mismatch and therefore bigger difference between the <111>$_\text{pc}$ directions of substrate materials compared to strained SCIO, in combination with peaks appearing already from the bare substrate material at the crucial $2\theta$ positions, is a key reason, why this access is denied in case of the other used substrate materials.

To clarify the B-site ordering state in such  SCIO films we performed the polarization-dependent Raman spectroscopy (Figure \ref{Fig_Bsite} (b)-(e)) and compare the data with other ordered double perovskites of the same point group, e.g. La$_2$CoMnO$_6$ (LCMO). B-site-ordered LCMO obeys a monoclinic $P12_1/n1$ structure, for which theoretical lattice dynamical calculations \cite{Iliev} predict $A_g$ stretching mode of the (Co/Mn)O$_6$ octahedra (breathing mode) at $\sim 697\,\text{cm}^{-1}$. In contrast, the disordered LCMO obeys an orthorhombic $Pbnm$ structure and possesses the $B_{1g}$ breathing mode. Similarly, we expect a (Co/Ir)O$_6$ breathing mode with $A_g$ symmetry at a Raman shift, $\text{RS}\sim 697\,\text{cm}^{-1}$, for the B-site-ordered SCIO. In this case the $A_g$ mode should be present in Raman spectra, measured in parallel PP- and P'P'-configuration, and it is forbidden in the crossed scattering PS- and P'S'-configuration\cite{Iliev,Truong,Meyer} (the prime i
 ndicates measurements with an in-plane rotation of the sample by 45$\,^\circ$ around the [001]$_\text{pc}$ direction and provide an additional tool to test the epitaxy). For the disordered orthorhombic structure, the selection rules are opposite\cite{Iliev,Truong,Meyer}. As one can see in Figure \ref{Fig_Bsite} (b)-(e), all films in our strain series exhibit a strong breathing mode at $\sim 697\,\text{cm}^{-1}$ in the parallel PP- and P'P'-scattering configurations and only a weak (or none) intensity in the crossed PS- and P'S'-configuration. Thus, the $A_g$ symmetry and the B-site ordering can be concluded for all strained films grown on GSO, DSO, LSAT and NGO. Furthermore, the Raman shift of the breathing mode (see Figure \ref{Fig_Bsite} (f)) was found to depend linearly on the strain, $\text{RS}\sim-a_\text{pc}^\text{substrate}$. This is due to the change of the unit cell volume, $V_\text{pc uc}$, and of the phonon energy with strain. Note, that a smaller $V_\text{pc uc}$
  leads to a closer packing and, respectively, more energy is necessary. Kumar and Kaur\cite{Kumar}, observed a similar behavior of the breathing mode in a strain relaxation series of the double perovskite La$_2$NiMnO$_6$/LaAlO$_3$(001).

For selected films on GSO, STO and NGO substrates the B-site ordering was also studied on the micro scale by HAADF STEM measurements along the [110]$_\text{pc}$ direction (see supplementary material) \cite{SM}. These measurements show that independent on strain there are regions with a high degree of B-site ordering, which alternate with regions with lower or no B-site ordering at all. Concluding, the epitaxial strain perpendicular to the [001]$_\text{pc}$ direction could neither improve nor weaken the B-site ordering in SCIO.

\subsection{Magnetic properties}

\begin{figure}[t]
\includegraphics[width=0.92\linewidth]{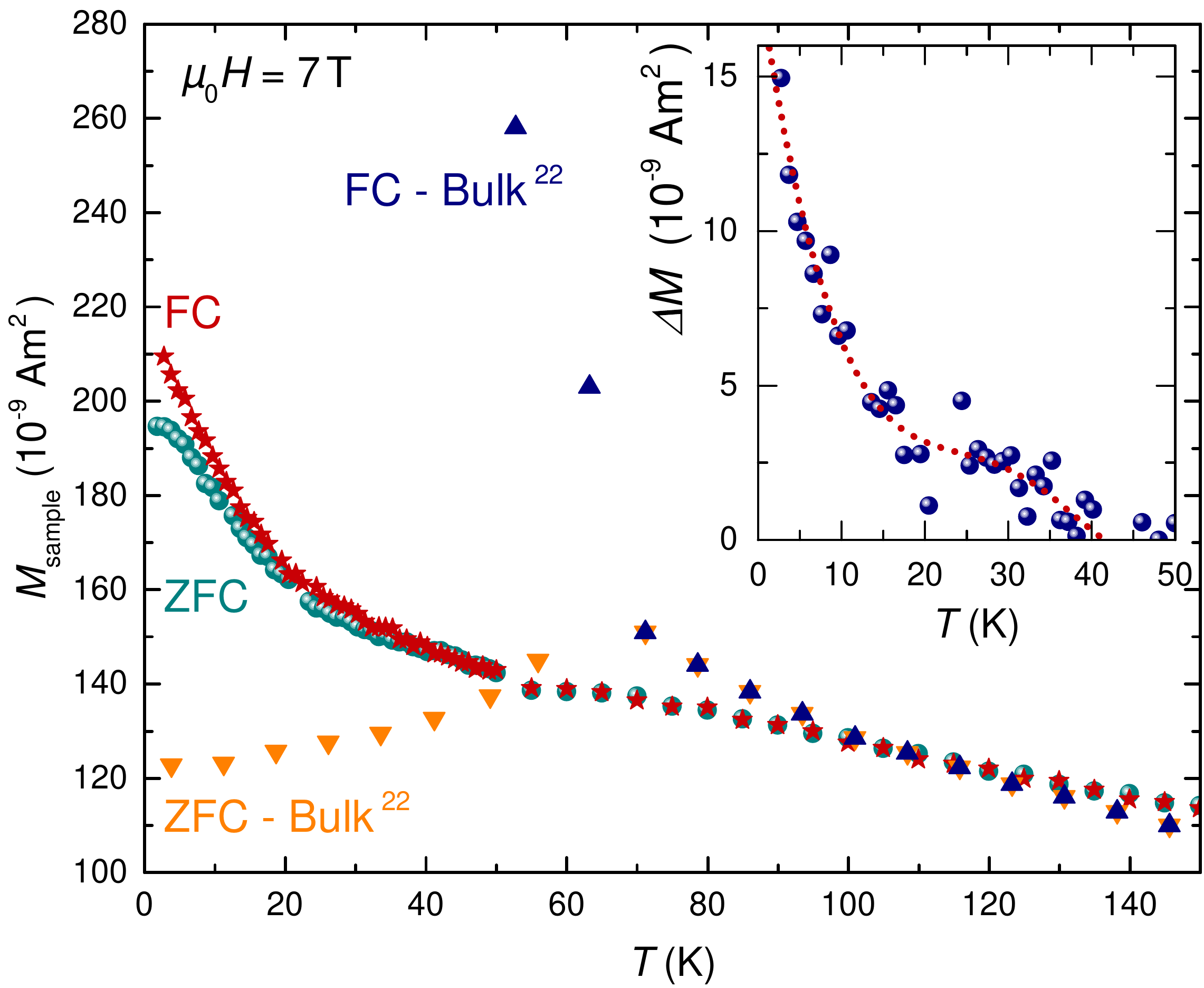}
\caption{\label{Fig_MPMS_MvsT}(Color online) Temperature dependence of FC and ZFC magnetization of $\sim 250\,\text{nm}$ thick (111)$_\text{pc}$ oriented SCIO (after subtraction of the substrate background contribution) in comparison to literature data of bulk polycrystalline SCIO from reference~\cite{Narayanan} (scaled to the sample volume). \textit{Inset}: Temperature dependence of the difference between the FC and ZFC data (dotted line intended to guide the eyes).}
\end{figure}

Separation of magnetic properties of thin films from the substrate magnetic contribution is generally difficult. A comparison with data on bulk SCIO~\cite{Narayanan} indicates, that a film with minimal thickness of $\sim$ 200\,nm is required to get magnetic moment corresponding to $\sim$ 5\,\% of the moment of pure STO substrate at room temperature. Because fully strained (001)$_\text{pc}$ oriented SCIO films are stable for d$\leq$50\,nm, we focus here solely on thick ($d\sim250\,\text{nm}$) strained SCIO/STO(111)$_c$ films, cf. Figure 1. However, we identified a Sr$_3$Co$_2$O$_6$ (SCO) second phase inclusion on the level 0.8\,volume-\%. For a quantitative determination of the SCO volume fraction, we have used the field dependent measurements of magnetization at lowest temperature (for more details, see supplemental material \cite{SM}). The foreign phase moments are saturated in a field of 7~T that has been applied in the measurement shown in Figure \ref{Fig_MPMS_MvsT}. Furth
 ermore, for our analysis of the difference between zero-field cooled (ZFC) and field cooled (FC) magnetization, it cancels out.

Previous investigation on polycrystalline SCIO revealed long-range magnetic order at 70~K, with ferromagnetic and antiferromagnetic components of the ordered magnetic moments, yielding to a pronounced difference in the FC and ZFC magnetization~\cite{Narayanan}. Figure \ref{Fig_MPMS_MvsT} compares these literature data on polycrystalline SCIO (from Ref. \cite{Narayanan}, indicated by triangles) with the FC and ZFC magnetization measurements of our thick (111)$_\text{c}$ oriented SCIO film below 150~K. Compared to the polycrystal data, much weaker splitting between FC and ZFC is found and the maximum at 70~K in ZFC mode is absent. Taking the onset of the difference $\Delta M$ between FC and ZFC (see inset of Figure \ref{Fig_MPMS_MvsT}) as measure of magnetic order, an ordering temperature of 43~K would be estimated for the film, which is significantly reduced compared to the polycrystal. We speculate that epitaxial strain and/or better B-site order is responsible for the differ
 ence in the magnetic susceptibility behavior.

In strained (111)$_\text{pc}$ oriented thin films of the double perowskite Sr$_2$FeMoO$_6$ Hauser \textit{et al.} found a similar decrease of the ordering temperature \cite{Hauser}.
Unfortunately, the thinner films of the (001)$_\text{pc}$ strain series could not be investigated by respective magnetization measurements and the evolution of the susceptibility behavior with strain remains therefore unknown.

\subsection{Spectroscopic results}

\begin{figure}
\includegraphics[width=0.95\linewidth]{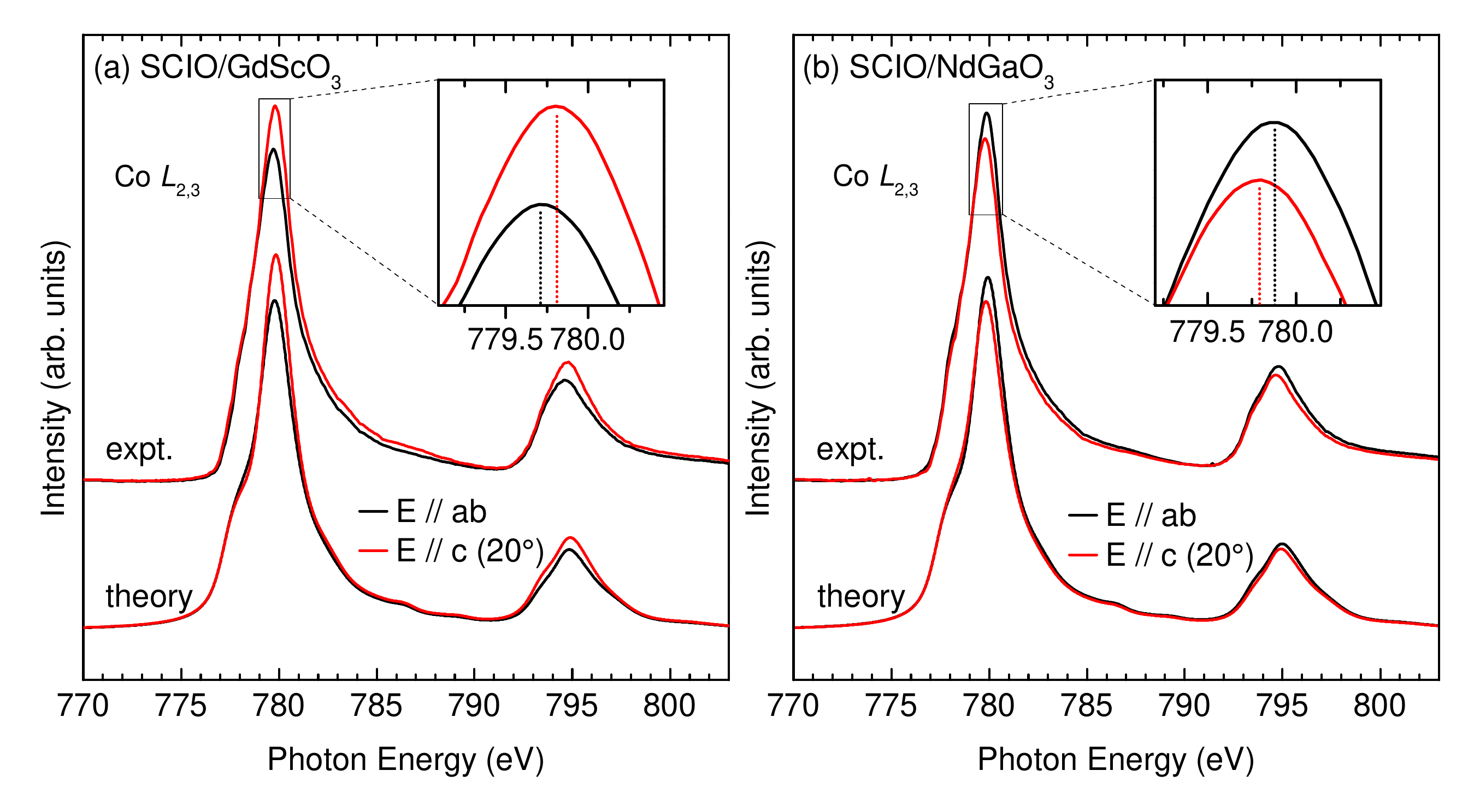}\newline
\includegraphics[width=1\linewidth]{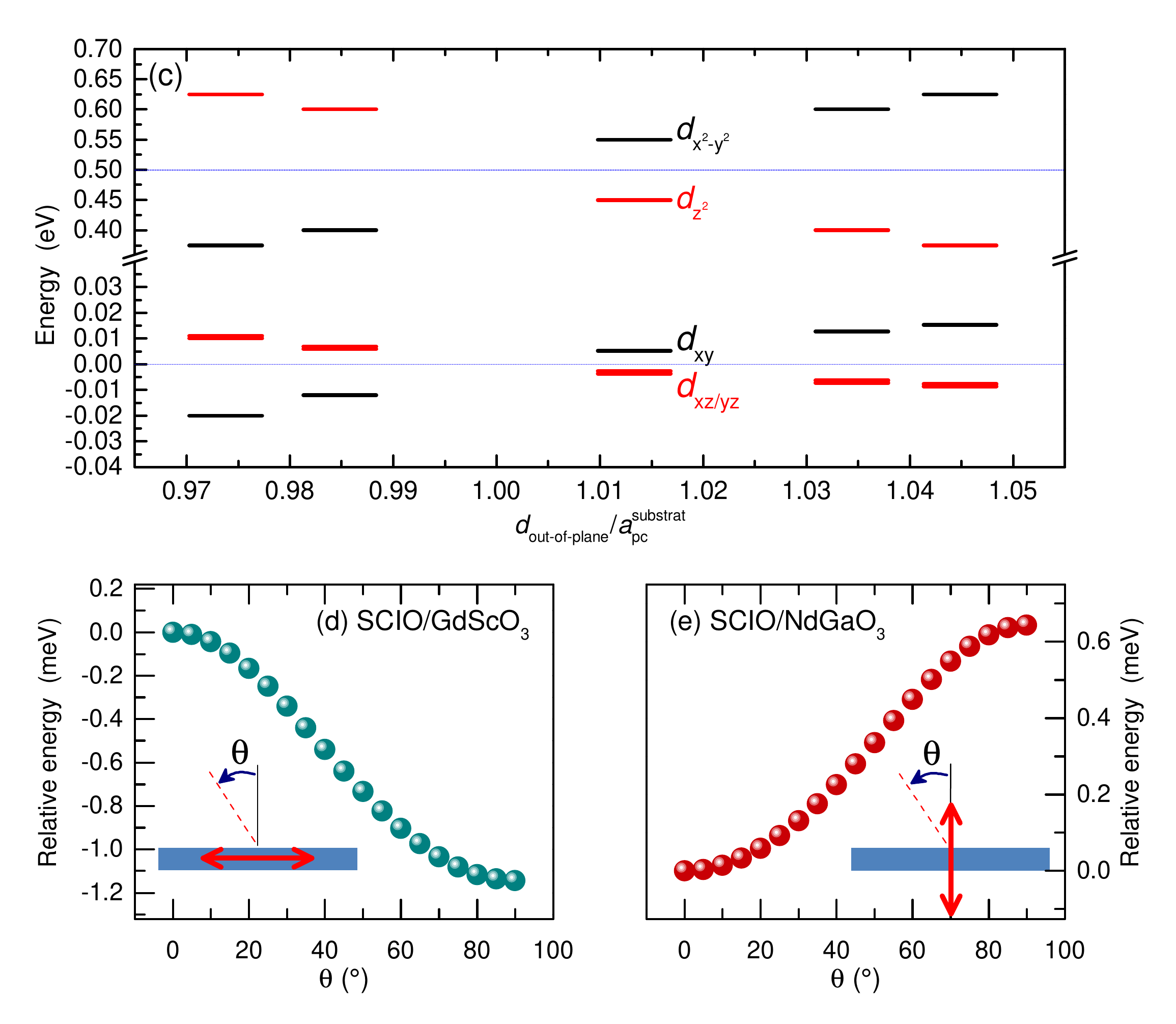}
\caption{\label{Fig_XAS}(Color online) Experimental polarization-dependent Co $L_{2,3}$ XAS spectra of (a) the most tensile strained SCIO films on GSO, and (b) the most compressive strained SCIO films on NGO in this study, taken at 300 K, together with the theoretical spectra calculated for Co$^{3+}$ high spin (HS) state. The inset of (a) and (b) is a zoom-in view at the $L_3$ main peak showing the favor orbital state under different strains. (c) The energy level diagram of the orbital states of Co$^{3+}$ in the SCIO films as a function of strain, obtained from the configuration interaction cluster model including the full atomic multiplet theory, which are consistent with the experimental polarization-dependent XAS results. (d) and (e) are the calculated magnetic anisotropy energy as a function of the spin direction of the SCIO films on GSO and NGO substrates, respectively. Each inset depicts a schematic drawing of the resulting AFM axis.}
\end{figure}

As mentioned in the previous section the magnetic signal for the samples of the (001)$_\text{pc}$ strain series are too weak to be detected in a conventional SQUID magnetometer. In order to investigate the electronic structure and its antiferromagnetic property of the SCIO thin films, we resort to polarization-dependent XAS. Optical measurements were done on the SCIO films capped by 2 nm thick STO~\cite{Esser}, which due to its high integrity and insulating properties~\cite{Belenchuk} protects SCIO films from degradation but still allows electrons to escape. We note that the polarization-dependent XAS is one of few techniques that can determine the magnetic axis of an antiferromagnetic ordered state in thin films. \cite{Haverkort2004,Csiszar2005,NatComms7} Figure \ref{Fig_XAS} (a) and (b) show the experimental polarization-dependent Co $L_{2,3}$ XAS spectra of the most tensile strained SCIO thin films on GSO and the most compressive strained SCIO films on NGO, respectively, in this study, taken at 300 K. The spectra of the SCIO thin films on DSO, STO, and LSAT substrates are shown in Figure S9 of the SM \cite{SM}. The spectra are dominated by the Co $2p$ core-hole spin-orbit coupling which splits the spectrum
  roughly in two parts, namely the $L_{3}$ ($h\nu \approx$ 776-784 eV) and $L_{2}$ ($h\nu \approx$ 793-797 eV) white lines regions. The line shape strongly depends on the multiplet structure given by the Co 3$d$-3$d$ and 2$p$-3$d$ Coulomb and exchange interactions, as well as by the local crystal fields and the hybridization with the O 2$p$ ligands. Unique to soft XAS is that the dipole selection rules are very sensitive in determining which of the 2$p^{5}$3$d^{n+1}$ final states can be reached and with what intensity, starting from a particular 2$p^{6}$3$d^{n}$ initial state ($n= 6$ for Co$^{3+}$) \cite{PRL92,PRL102,PRB95}. This makes the technique extremely sensitive to the symmetry of the initial state, i.e., the spin, orbital and valence states of the ions.

The experimental XAS spectra show that the Co valence state of these SCIO thin films is mainly 3+.
Please be noted that for comparing with the calculated spectra the experimental spectra in Figure \ref{Fig_XAS} (a) and (b) have been subtracted by 10\% and 8\% of Co$^{2+}$, respectively. The experimental isotropic XAS spectra before subtracting Co$^{2+}$ contributions can be found in Figure S8 of the SM \cite{SM}. Further, the spectral features indicate a high spin (HS) state of the Co$^{3+}$ ions \cite{PRL92,PRL102,PRB95} in the SCIO thin films which is independent of substrates underneath, i.e., a strain independent HS state. On the other hand, the orbital state in these SCIO thin films is quite different as indicated by the opposite sign of the polarization-dependent difference of spectra.\cite{Haverkort2004,Csiszar2005} For example, the intensity of the $L_3$ main peak is always larger for E $\parallel$ c (20$^{\circ})$ than for E $\parallel$ ab in the SCIO/GSO thin film, whereas it is always smaller in the SCIO/NGO thin film. Since these spectra were taken at 300 K whi
 ch is much higher than the magnetic ordering temperature of 43\,K, the polarization contrast is caused solely by crystal field effects. Using the E vector of light parallel to the ab plane (black lines), we can reach the unoccupied Co 3$d$ orbital states with xy/x$^2$-y$^2$ characters. With the E vector of light parallel c axis (red lines), we detect the unoccupied Co 3$d$ orbital states with yz/zx/3z$^2$-r$^2$ characters. This indicates that the sign of the crystal field splitting is opposite in the two systems. This inference is further consolidated by the observed peak energy difference of the $L_3$ main line as depicted in the insets of Figure~\ref{Fig_XAS} (a) and (b). The $L_3$ peak position is higher for E $\parallel$ c (20$^{\circ})$ than for E $\parallel$ ab in the SCIO/GSO thin film, while it is lower in the SCIO/NGO thin film. All together, we derive that in the tensile strained SCIO/GSO thin films the orbital states with x/y-character are energetically favorable,
  whereas in the compressive strained SCIO/NGO thin films the orbital states with z-character are energetically favorable. This can be understood qualitatively as the applied strain from the substrates underneath induces a tetragonal distortion and causes the corresponding orbital state shift in energy.

For further confirming this orbital energy level diagram and knowing the corresponding magnetic anisotropy, we have simulated the XAS spectra using the well proven configuration interaction cluster model that includes the full atomic multiplet theory \cite{deGroot,Tanaka}. The calculations were performed using the XTLS 8.3 program \cite{Tanaka}. For the calculation details, please see the SM \cite{SM}. As displayed in the bottom of Figure \ref{Fig_XAS} (a) and (b), the calculated spectra based on the HS Co$^{3+}$ model with the energy diagram of the orbital state shown in Fig. 9 (c) can well reproduce the experimental spectra. We can safely conclude that a tensile strain stabilizes a ($3d^{5}\uparrow\uparrow\uparrow\uparrow\uparrow$ + 3$d_{xy}\downarrow$) state and a compressive stain stabilizes a ($3d^{5}\uparrow\uparrow\uparrow\uparrow\uparrow$ + 3$d_{yz/zx}\downarrow$) state in SCIO thin films. We infer that the change of the anisotropy in the crystal field parameters in going from room temperature to low temperatures due to thermal contraction of the substrates is negligible, and that we therefore can use these XAS derived parameters also for the analysis of the low temperature magnetic properties (see the Supplemental Material for details). Accordingly, we can calculate the magnetic anisotropy energy as a function of the spin direction for each case as shown in Figure \ref{Fig_XAS} (d) (tensile, SCIO/GSO) and (e) (compressive, SCIO/NGO).The magnitude of the exchange field was set to 4 meV in accordance with the the magnetic ordering temperature of 43 K. \cite{PRB95} This magnetic anisotropy energy is expressed as $E = K_{0} + K_{1}\sin^{2}(\theta) + K_{2}\sin^{4}(\theta) + K_{3}\sin^{4}(\theta)\sin^{2}(\varphi)\cos^{2}(\varphi)$, where $\theta$ is the angle between the exchange field and the [001]$_{pc}$, and $\varphi$ is the azimuthal angle which is set to 45$^{\circ}$. We find for the SCIO/GSO (tensile) $K_{1} = -1.42$ meV, $K_{2} = 0.14$ meV, and $K_{3} = 0.56$ meV, while for the SCIO/NGO (compressive) we obtain $K_{1} = 0.48$ meV, $K_{2} = 0.08$ meV, and $K_{3} = 0.33$ meV. In other words, for the tensile strained SCIO/GSO thin film, the spin moment favors the in-plane direction with the energy difference of about 1.14 meV between the magnetic easy axis ($\theta = 90^{\circ}$, in the 
 film plane) and hard axis ($\theta = 0^{\circ}$, perpendicular to the film), whereas for the compressive strained SCIO/NGO thin film, the spin moment favors the out-of-plane direction with the energy difference of about 0.64 meV between the magnetic easy axis ($\theta = 0^{\circ}$, perpendicular to the film) and hard axis ($\theta = 90^{\circ}$, in the film plane). Note that the evaluated strain-induced AFM magnetic anisotropy in SCIO films, i.e. in/out of plane for tensile/com-pressive stress, differs from that observed for ferromagnetic double perovskite films of La$_2$CoMnO$_6$, i.e. in/out of plane for compressive/tensile stress~\cite{Galceran,Lopez-Mir}. The reason is unclear up to know and, likely, is related to the FM exchange interaction between Co$^{2+}$ and Mn$^{4+}$ ions accordiung to the second Goodenough-Kanamori-Anderson rule. This is a piece of useful information for understanding the magnetotransport properties of the SCIO thin films under different strains, see the discussion in the section of Magnetotransport.

\subsection{Zero field transport properties}

\begin{figure}[t]
\includegraphics[width=0.95\linewidth]{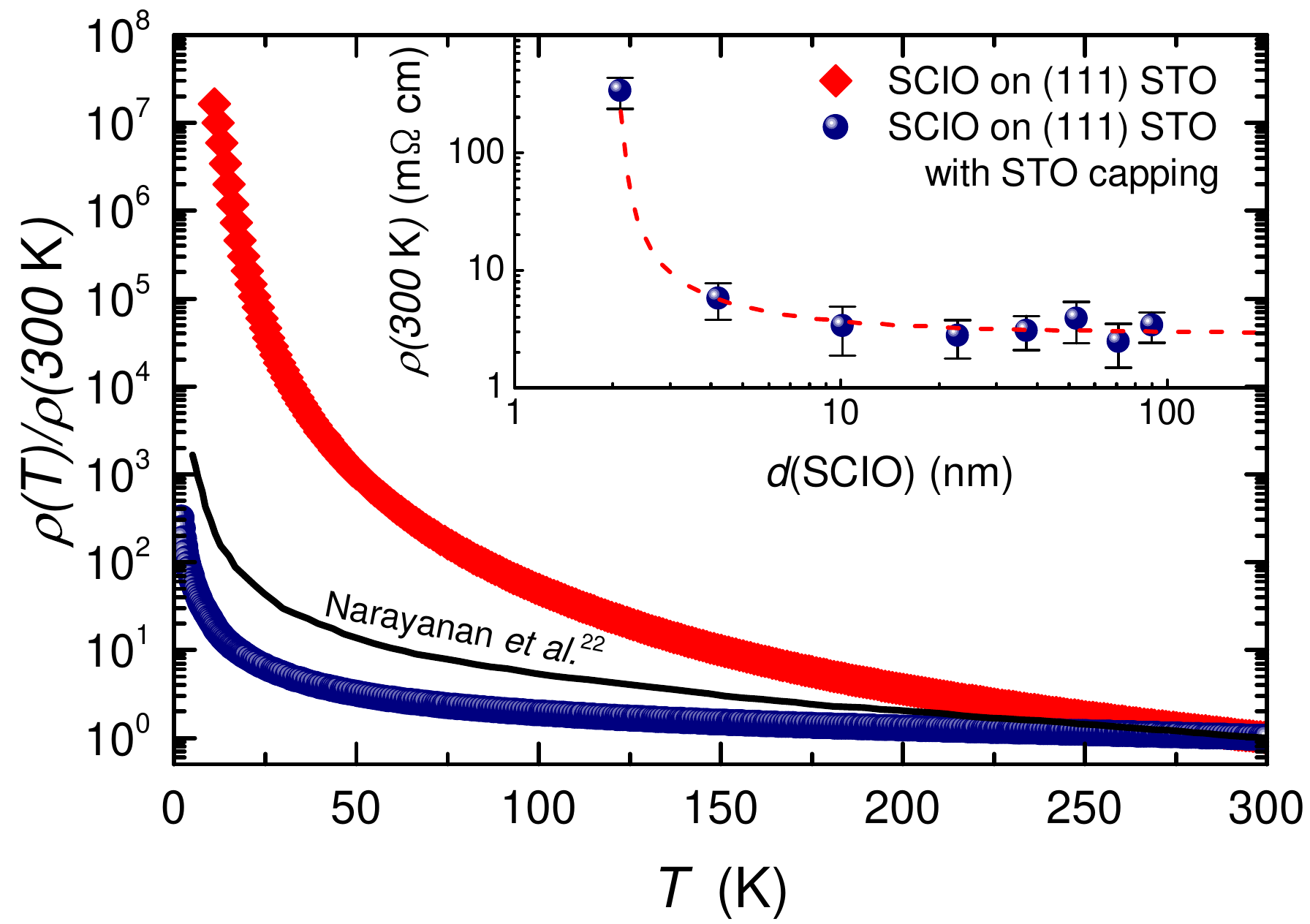}
\caption{\label{Fig_RT-111}(Color online) Temperature dependence of the electrical resistivity of SCIO for a sample with STO protection (blue) and without protection (red) in comparision to bulk data (black line) obtained by Narayanan \textit{et al}. \cite{Narayanan}. \textit{Inset}: Thickness dependence of specific resistance at 300\,K (dashed line as guide for eyes).}
\end{figure}

We now turn to the electrical resistivity of SCIO thin films. The temperature dependent $\rho(T)$ measurements for a (111)$_\text{pc}$ oriented $d$(SCIO) $=$ 30\,nm thick SCIO film (with and without STO protective top layer) are shown in Figure \ref{Fig_RT-111}. Similar to the bulk SCIO (Ref. \cite{Narayanan}), thin films also show an insulating behavior. The values of room temperature resistivity for the air protected and unprotected films were $ 3.08(1)\,\text{m}\Omega\,\text{cm}$ and $6.66(1)\,\text{m}\Omega\,\text{cm}$, respectively. The reported bulk value (Ref. \cite{Narayanan}) was about 3 and 1.5 times larger, respectively. Upon cooling to 4~K, the electrical resistance of the unprotected film  strongly increases, becoming several orders of magnitude larger than the resistance of the protectedfilm; the $\rho(T)$ of the bulk SCIO lies in between. This indicates the importance of the STO capping layer to protect the (surface) properties of SCIO. As shown in the inset of
  Figure \ref{Fig_RT-111} the room temperature resistivity is almost independent on the film thickness for $d>5-8\,\text{nm}$. Unfortunately, it was impossible to probe the thickness dependence at low temperatures.
Fitting the $\rho(T)$ data of the capped SCIO thin film between 250 and 300~K results in a charge gap of $32(5)$~meV \cite{SM}. Below 70~K, $\rho(T)$ indicates variable range hopping behavior \cite{SM}.

The electrical resistivity of the various films, strained along the (001)$_\text{pc}$ direction, is discussed in supplemental material \cite{SM}. The overall insulating behavior does not change, though a systematic enhancement of $\rho$(300 K) with strain is observed \cite{SM}.

\subsection{Magnetotransport within the (001)$_\text{\textbf{pc}}$ strain series}

\begin{figure}[b]
\includegraphics[width=0.95\linewidth]{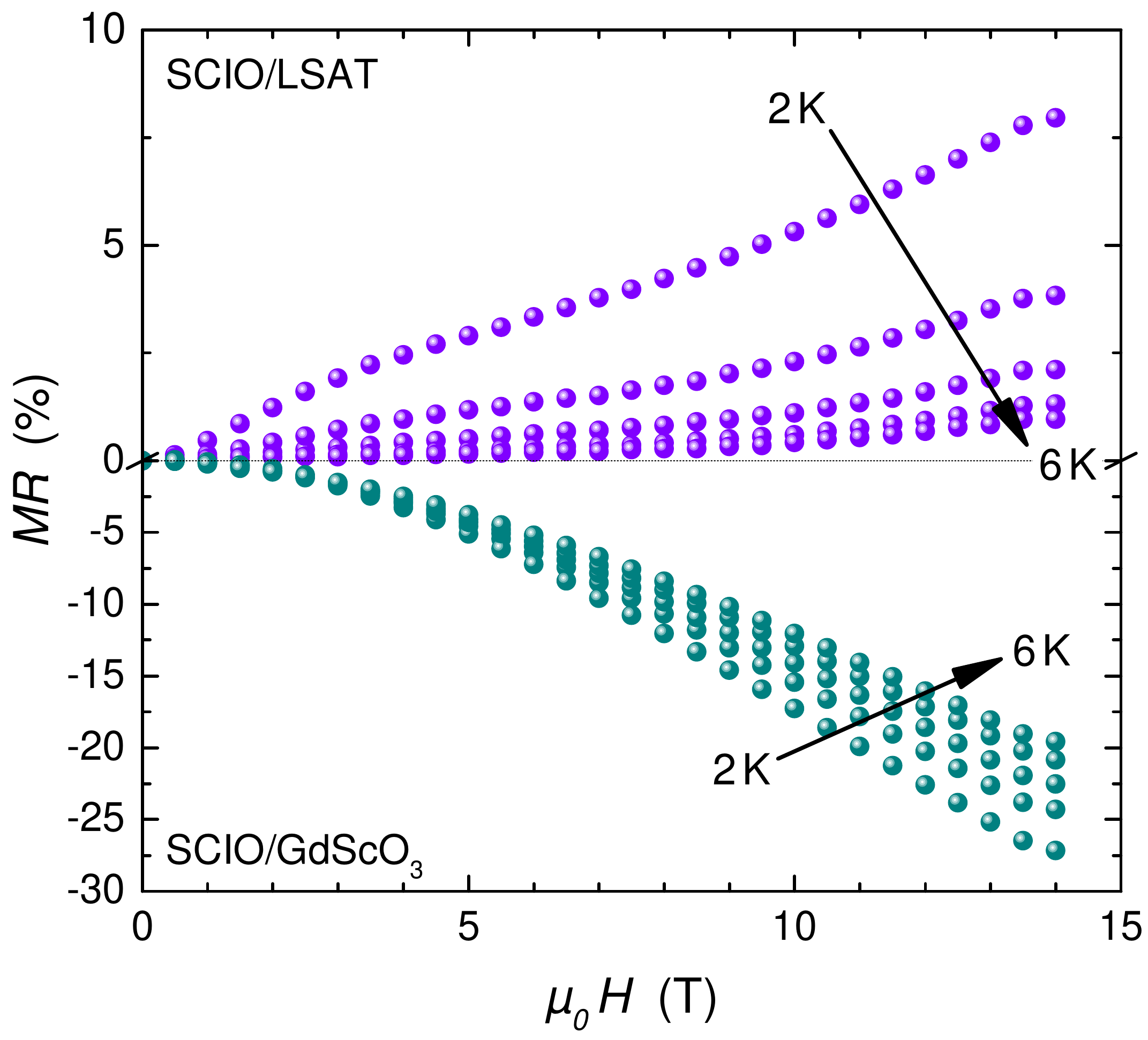}
\caption{\label{Fig_MR-overview} Isothermal transverse magnetoresistance $MR = (\rho(H)-\rho(0))/\rho(0)$ of (001)$_\text{\textbf{pc}}$ strained SCIO thin films grown on LSAT (compressive strain) and GdScO$_3$ (tensile strain) at various temperatures between 2 and 6 K (fields always applied transverse to the film planes).}
\end{figure}

\begin{figure}[t]
\includegraphics[width=0.95\linewidth]{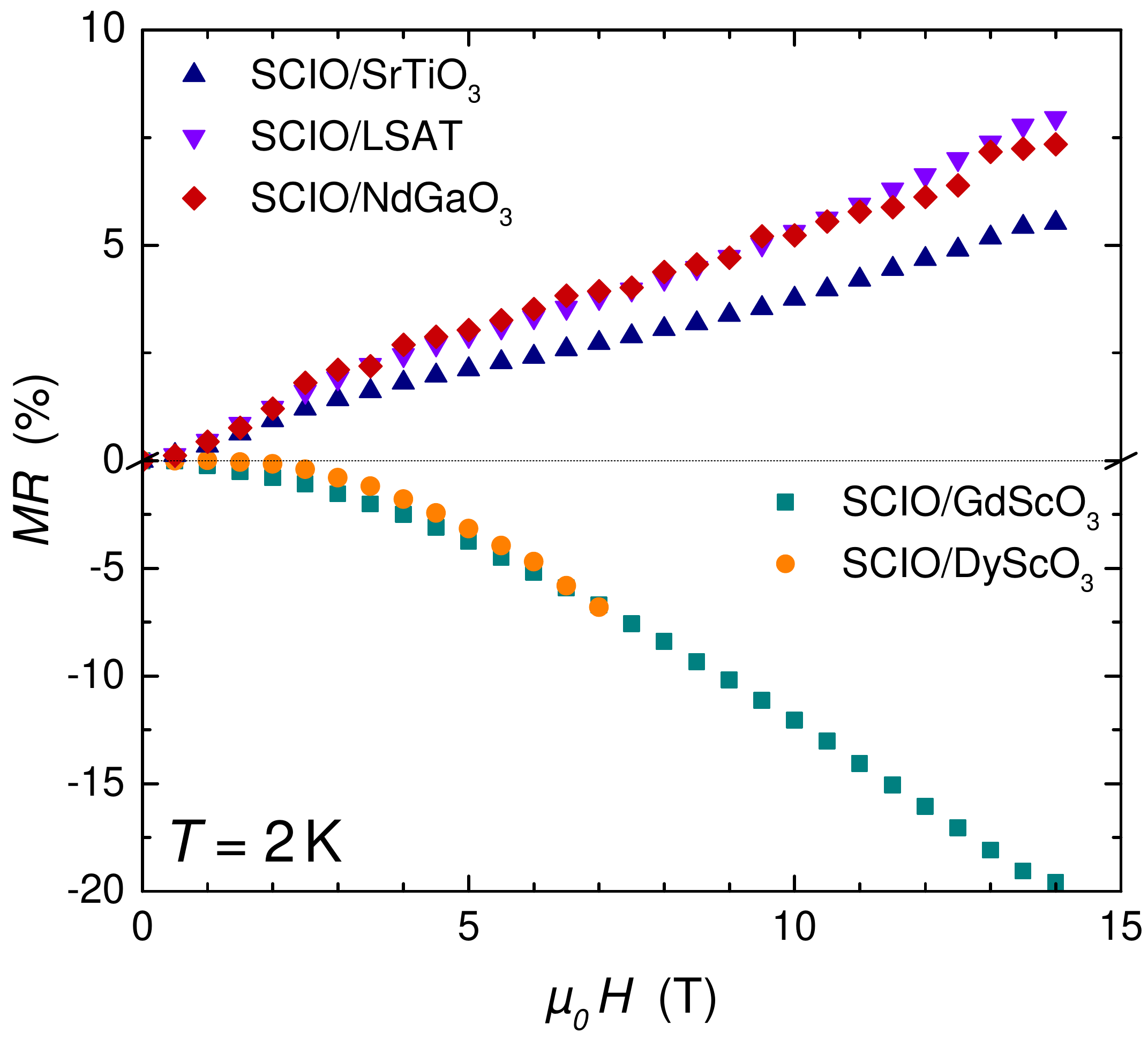}
\caption{\label{Fig_MR-strain-overview} Isothermal transverse magnetoresistance of all investigated (001)$_\text{\textbf{pc}}$ strained SCIO thin films at $T=2\,\text{K}$. Note, that the thin film on DSO substrate broke in a field of $\mu_0H=7\,\text{T}$ due to a high magnetic torque effect from DSO.}
\end{figure}

Considering the information on the magnetic anisotropy, obtained in the ''Spectroscopy'' section, we can now address the magnetotransport properties of the SCIO thin films under different strains.
Next, we focus on the low-temperature isothermal magnetoresistance (MR) of (001)$_\text{\textbf{pc}}$ strained SCIO thin films. As discussed previously in section C, the magnetic easy axis in this series changes from the ''in-plane'' to the ''out-of-plane'' as the strain changes from tensile to compressive. This change of the magnetic anisotropy has a direct influence on the magnetoresistance. As shown in Figure \ref{Fig_MR-overview}, MR below 6~K is {\it positive} for thin films grown under compressive strain on LSAT substrate and {\it negative} for those under tensile strain on GSO substrate. This indicates a direct relation between MR sign and magnetic anisotropy. This is further corroborated by the comparison of the MR at 2~K of all strained SCIO thin films, which, as shown in Figure \ref{Fig_MR-strain-overview}, follow this trend. This observation suggests a strong influence of the anisotropic magnetoresistance (AMR) effect in combination with the reorientation of the AF
 M ordered Co sublattice in SCIO in an external transverse magnetic field. 
\begin{figure}[b]
\includegraphics[width=0.5\linewidth]{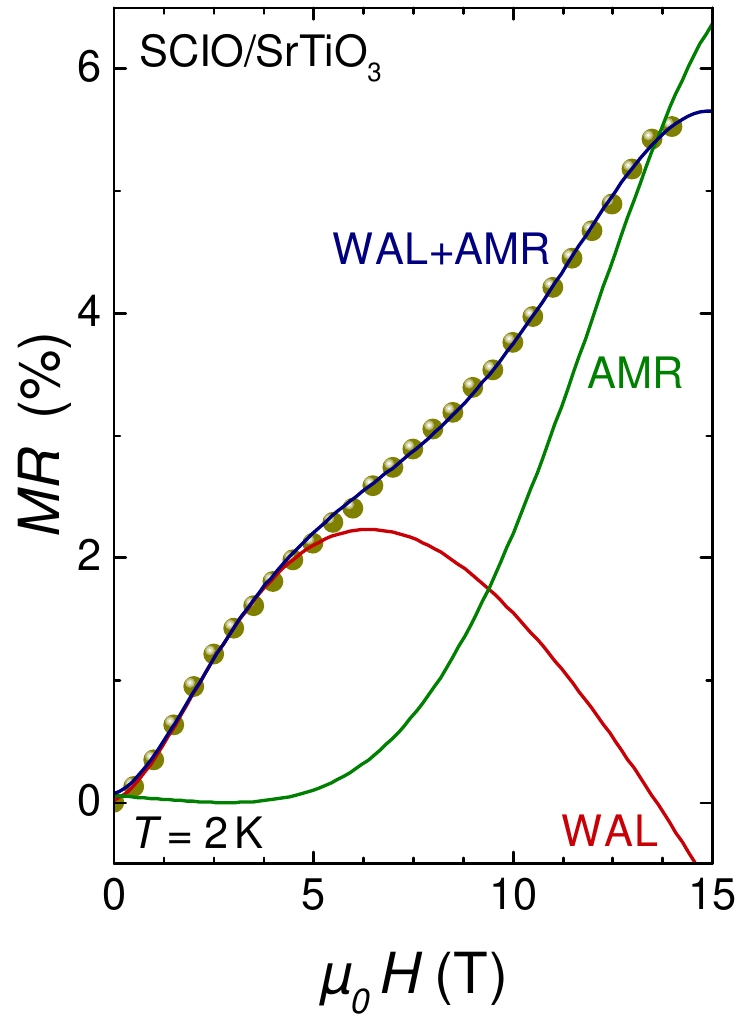}
\caption{\label{Fig_MR-STO}Field dependence of the MR of SCIO on STO at $T=2\,\text{K}$. The red and green lines indicate the contributions of the WAL and AMR effect to the total signal. For the latter contribution, a smeared spin flip scenario, detailed in supplemental materials \cite{SM}, has been utilized.}
\end{figure}
In addition, structural disorder in combination with large spin orbit coupling will generate a quantum correction to MR due to weak anti-localization (WAL). Based on the Dresselhaus effekt~\cite{Dresselhaus} we used for the description of the WAL term of the MR a Dresselhaus like contribution with isotropic spin orbit scattering, see supplemental material~\cite{SM}. For the AMR contribution the orientation of the moments of Co sublattices in their AF ground state with respect to the field direction is important. For tensile strain at zero field the moments are oriented in the film plane and application of a transverse field leads to a continuous rotation of both sublattice moments out of the plane. This leads to a negative MR. For compressive strain, the moments of the two sublattices initially point parallel and antiparallel to the applied field. If there would be very weak coupling between the two sublattices, only the moments of the antiparallel sublattice would continuous
 ly rotate towards the applied field direction with increasing field. In the alternative case, upon increasing magnetic field the moments remain in their orientation until a spin-flip occurs (for a sketch, see supplemental material), followed by continuous rotation towards the field. Since there is no indication for a sharp spin flip, we modeled the MR behavior in a smeared spin flip scenario. For more details, we refer to supplemental material~\cite{SM}.

By combining AMR and WAL a valuable quantitative description of the measured MR is possible for all strained SCIO thin films. This is exemplified in
Figure \ref{Fig_MR-STO} for compressively strained SCIO. Within the smeared spin-flip scenario, a positive AMR contribution to the MR dominates between 10 and 24~T~\cite{SM}. As detailed in supplemental material, the obtained parameter for the WAL and AMR contributions within the smeared spin-flip scenario are more realistic compared to the assumption of very weak coupling between the two Co sublattices. To point out our key arguments: We expected a positive contribution from AMR and a temperature independent spin-orbit scattering field within the WAL contribution. Both is only the case in the smeared spin-flip scenario.

\begin{figure}[t]
\includegraphics[width=0.95\linewidth]{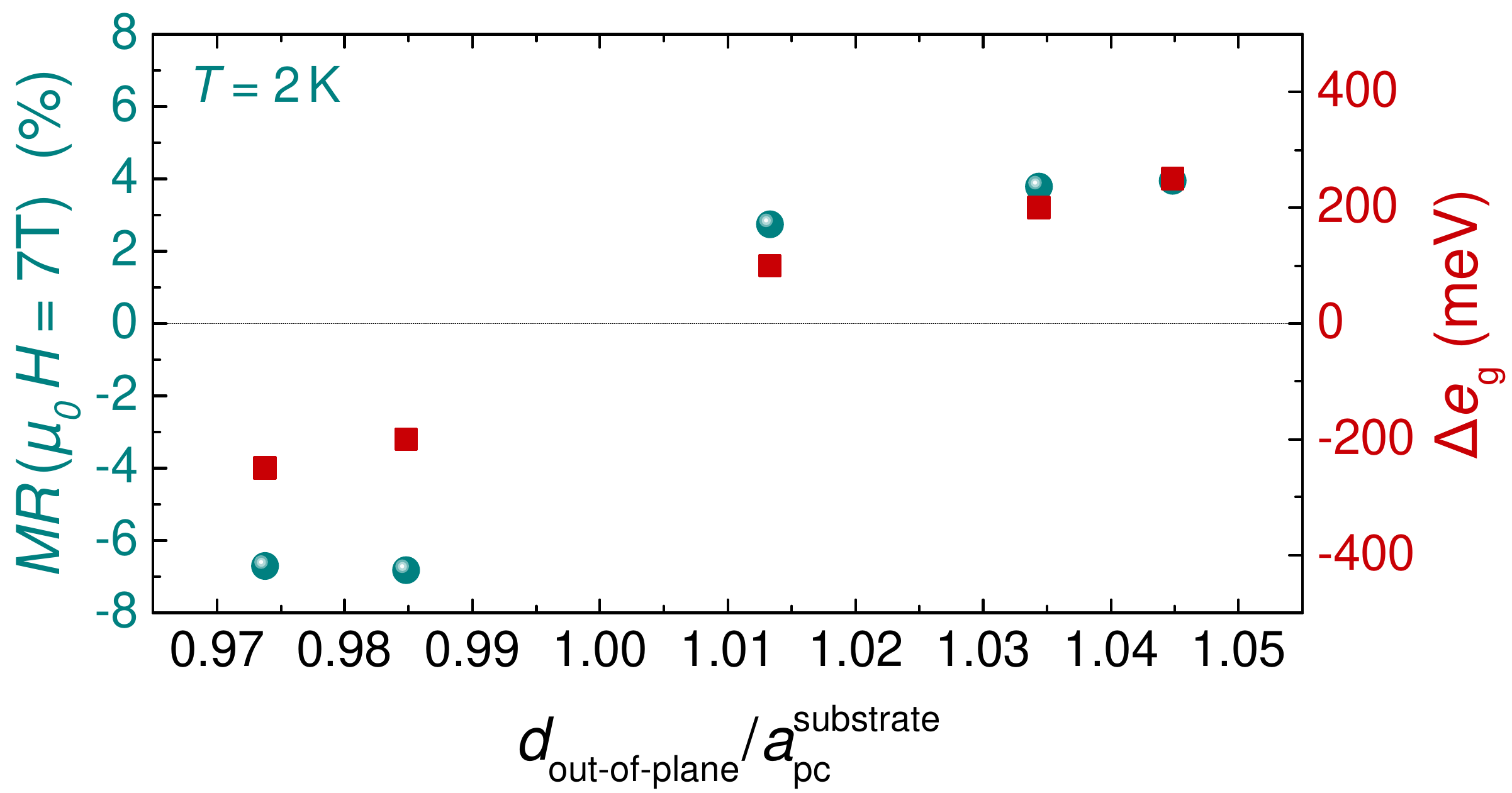}
\caption{\label{Fig_MR-strain}(Color online) Strain dependence of the MR at a magnetic field of $\mu_0H=7\,\text{T}$ at $T=2\,\text{K}$ (cyan circles) in comparison to the energy difference $\Delta e_\text{g} = E_{d_{x^2-y^2}}-E_{d_{z^2}}$ (red squares) from Figure \ref{Fig_XAS} (c).}
\end{figure}

As indicated by Figure \ref{Fig_MR-strain} the strain induced change of the Co d-level splitting, and respective change of moment orientation from in-plane (for tensile strain) to out-of plane (for compressive strain) goes hand in hand with a change of the MR from negative to positive. Due to the counteracting contributions of WAL and AMR effect in the samples with compressive strain, the absolute value of the MR is much smaller than for tensile strained samples, cf. Figure \ref{Fig_MR-strain-overview}.

\section{Conclusion}
For the first time B--site ordered SCIO thin films have been grown on (111)$_\text{c}$ oriented STO substrates and within a strain series also on various (pseudo) cubic (001)$_\text{pc}$ oriented substrates. Our electrical transport measurements of (111)$_\text{c}$ oriented samples with and without an air protection layer out of STO indicated in comparison to literature results from reference \cite{Narayanan} that the SCIO thin films degenerate in direct air contact. This leads us to the development of a four step \textit{in-situ} lithographical process to investigate the temperature and magnetic field dependencies of the electrical transport properties for protected samples.

The complete strain transfer from the substrate to the thin film was checked in all cases of the (001)$_\text{pc}$ strain series by RSMs around the (013)$_\text{pc}$ substrate peak in combination with HAADF STEM images in [100]$_\text{pc}$ direction. The, at least partial, B--site ordering of the SCIO thin films was proven by XRD scans in tilted geometry as well as by polarization dependent Raman spectroscopy. HAADF STEM images in [110]$_\text{pc}$ direction of selected samples reinforce these results.

The magnetic properties were investigated with SQUID magnetometry. The AFM ordering temperature $T_\text{N}\sim 43(10)\, \text{K}$ of the thick SCIO thin films is reduced in comparison to the bulk value ($T_\text{N}\sim 70\, \text{K}$ \cite{Narayanan}).
Using polarization dependent x-ray absorption spectroscopy at the Co $L_{2,3}$ edges, we revealed that the Co $3d$ orbital occupation strongly depends on the strain in the SCIO thin films induced by the substrates. The tensile strained SCIO thin films stabilize the occupation of the minority orbital state with x/y character, whereas the compressive strained SCIO thin films favor the occupation of the minority orbital state with z character. Together with the calculations using the well proven configuration interaction cluster model, we were able to determine the magnetic easy axis change due to the induced strain from the substrates, and the sign and the magnitude of the magnetic anisotropy energy of the antiferromagnetically ordered high-spin Co$^{3+}$ ions. We presented a quantitative model including the anisotropic magnetoresistance and weak anti-localization effects to explain the opposite behavior and its magnitude of the low-temperature magnetoresistance in the tensile 
 strained from the compressive strained SCIO thin films.

\begin{acknowledgments}
The authors like to thank S. Meir for developing the shadow mask argon ion etching method at Augsburg University and B. Meir for the support with it. We thank R. Pentcheva and P. Seiler for stimulating discussions. The work at Augsburg was supported by the Deutsche Forschungsgemeinschaft (DFG) through SPP 1666 and TRR 80. The research in Dresden was partially supported by the DFG through SFB 1143 and the study in G{\"o}ttingen through SFB 1073 (TP A02).
\end{acknowledgments}

\clearpage


\maketitle
\makeatletter
\renewcommand{\thefigure}{S\@arabic\c@figure}
\makeatother
\setcounter{figure}{0}
\section*{Supplemental Material}
In this Supplemental Material we provide a detailed description of the \textit{in-situ} lithographical process, additional HAADF STEM images in [110]$_\text{pc}$ direction to clarify the B--site ordering and x-ray characterization as well as the field dependence of the magnetization of the sample measured in the conventional SQUID magnetometer to estimate the volume fraction of the magnetic impurity phase. Furthermore more details are given on the electronic transport properties, including a detailed evaluation of the field dependence of the magnetoresistance by combining weak anti-localization and AMR effect, and details to the XAS spectra and the configuration interaction cluster calculation.

\subsection*{\textit{In-situ} lithographical process}
\begin{figure}[b]
\includegraphics[width=0.95\linewidth]{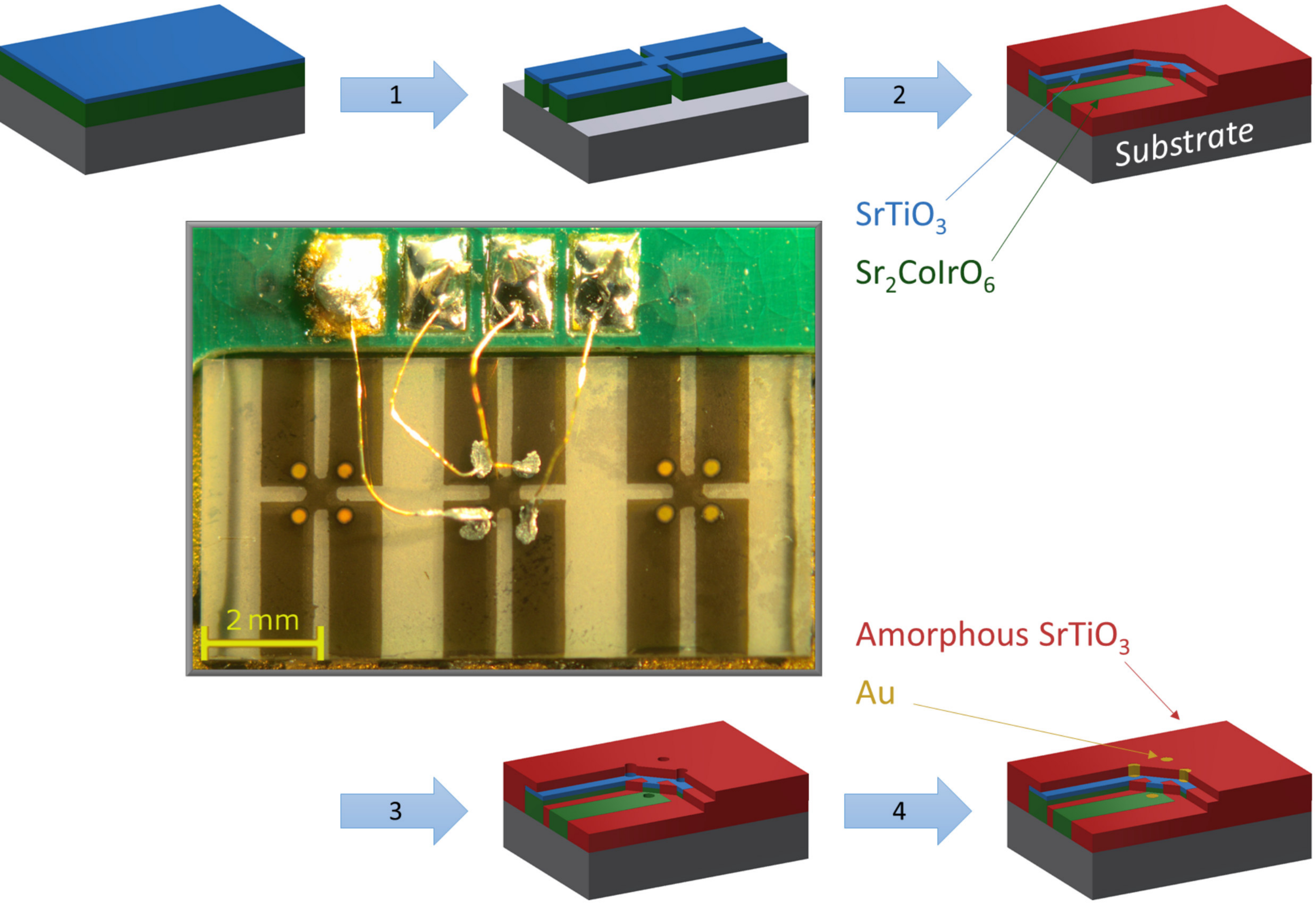}
\caption{\label{Fig_Litho} Schematic sketch of the \textit{in-situ} lithographical process and a photo of a processed and mounted STO capped SCIO sample. Gold wires ensure a good connection to the PPMS puck and were glued with silver paste to the gold contact pads on the sample.}
\end{figure}
As mentioned in the main text we developed a four step method (sketched in Figure \ref{Fig_Litho}) with the following procedure to exclude an air exposure of the samples during the lithographic process:
\begin{enumerate}
	\item A van der Pauw measurement geometry is structured by using a shadow mask argon ion etching technique.
	
	\item The shadow mask is removed and for air protection a thick amorphous STO layer is ablated by pulsed laser deposition using a KrF (248\,nm) excimer laser focused into a vacuum chamber to an energy density of 2-3 J/cm$^2$ at the target at room temperature with an O$_2$ background gas pressure of $0.5\,\text{mBar}$.
	
	\item A second shadow mask argon ion etching process digs contact holes through the STO cappings into the SCIO layer at the contact pads of the Van der Pauw measurement geometry.
	
	\item Those holes are filled with RF sputtered gold through the holes of the second shadow mask.
\end{enumerate}
A structured and contacted sample is shown in the photo of Figure \ref{Fig_Litho}.

\subsection*{HAADF STEM measurements in [110]$_\text{\textbf{pc}}$ direction}

\begin{figure}[b]
\includegraphics[width=0.95\linewidth]{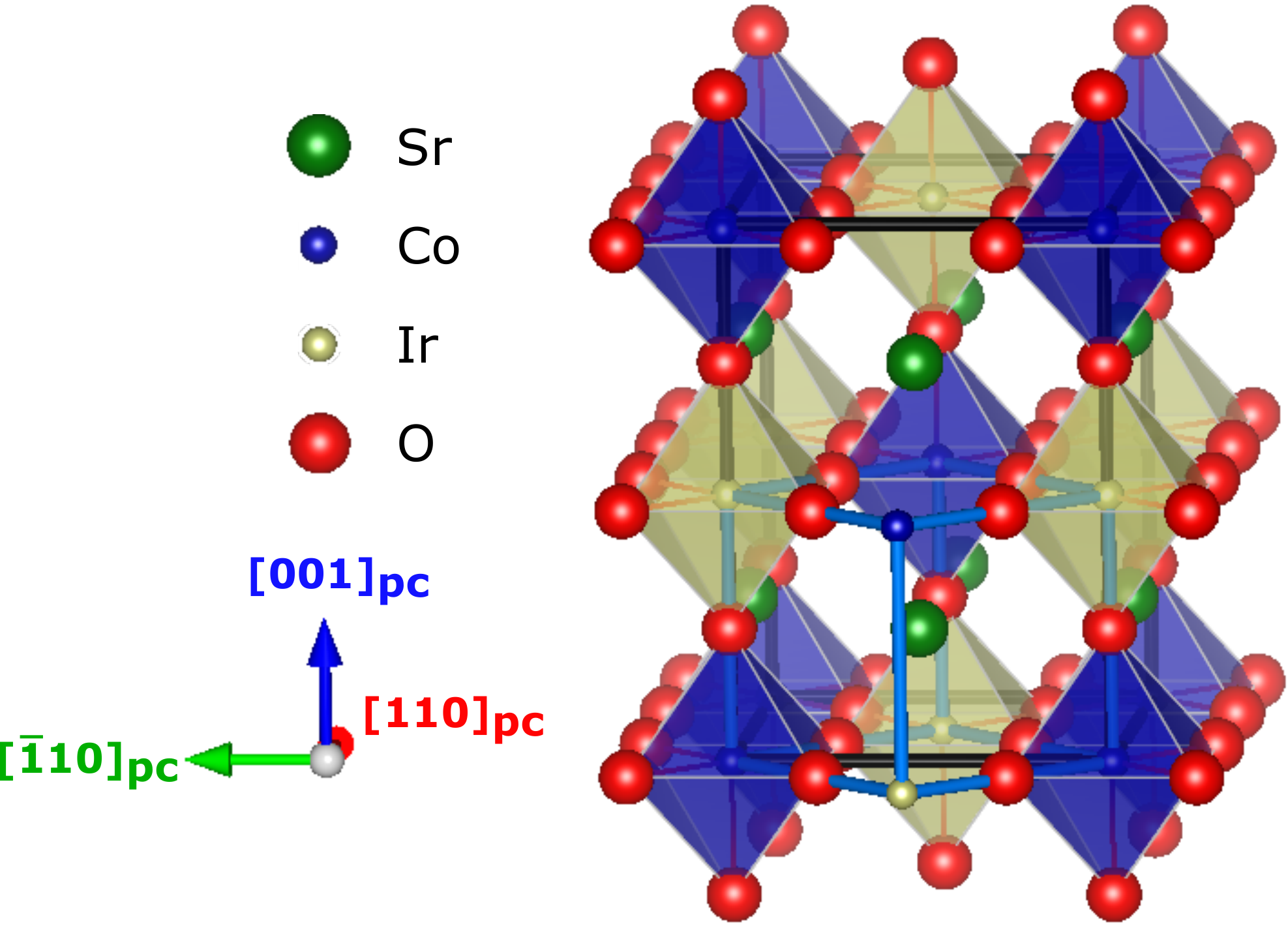}
\caption{\label{Fig_Crystal_110}Crystal structure of SCIO in [110]$_\text{pc}$ direction visualized with VESTA \cite{S_VESTA} based on crystal data from reference \cite{S_Mikhailova}.}
\end{figure}

\begin{figure*}[t]
\includegraphics[width=0.95\linewidth]{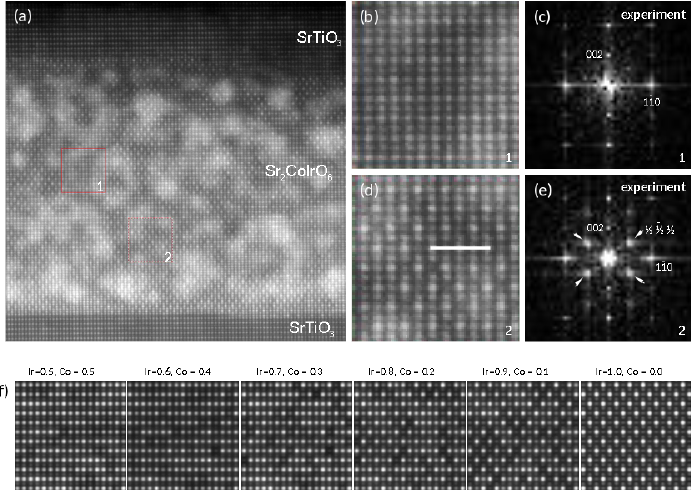}
\caption{\label{Fig_TEM_110_NGO} (a) Experimental HR-STEM image; (b) Enlarged area 1 from (a), and its FFT (c); (d) Enlarged area 2 from (a), and its FFT (e); (f) Simulated HRSTEM images. Occupancies are given for the Ir in positions (0,0,0) and (0.5, 0.5, 0.5), Co in positions (0.5, 0.5, 0) and (0, 0.5, 0.5), respectively. The best fit between the simulated and experimental images  is for Ir=0.8-0.9 and Co = 0.2-0.1 content, respectively.}
\end{figure*}

\begin{figure}[h!]
\includegraphics[width=0.95\linewidth]{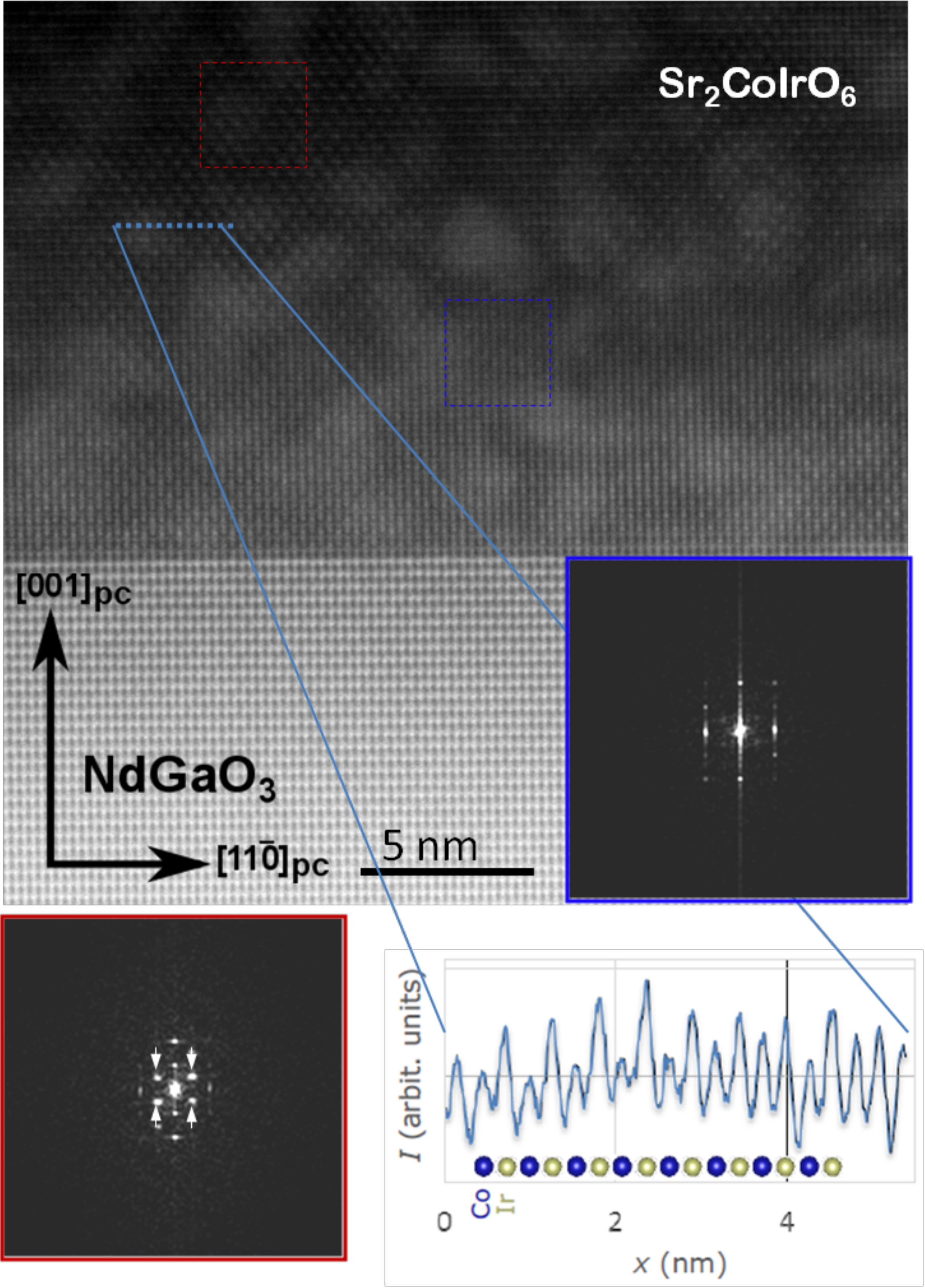}
\caption{\label{Fig_TEM_110_STO}HAADF STEM image in [110]$_\text{pc}$ direction of a SCIO thin film on (001)$_\text{c}$ oriented NGO including a line scan proving the alternating Co/Ir arrangement.}
\end{figure}

\begin{figure}[h!]
\includegraphics[width=0.9\linewidth]{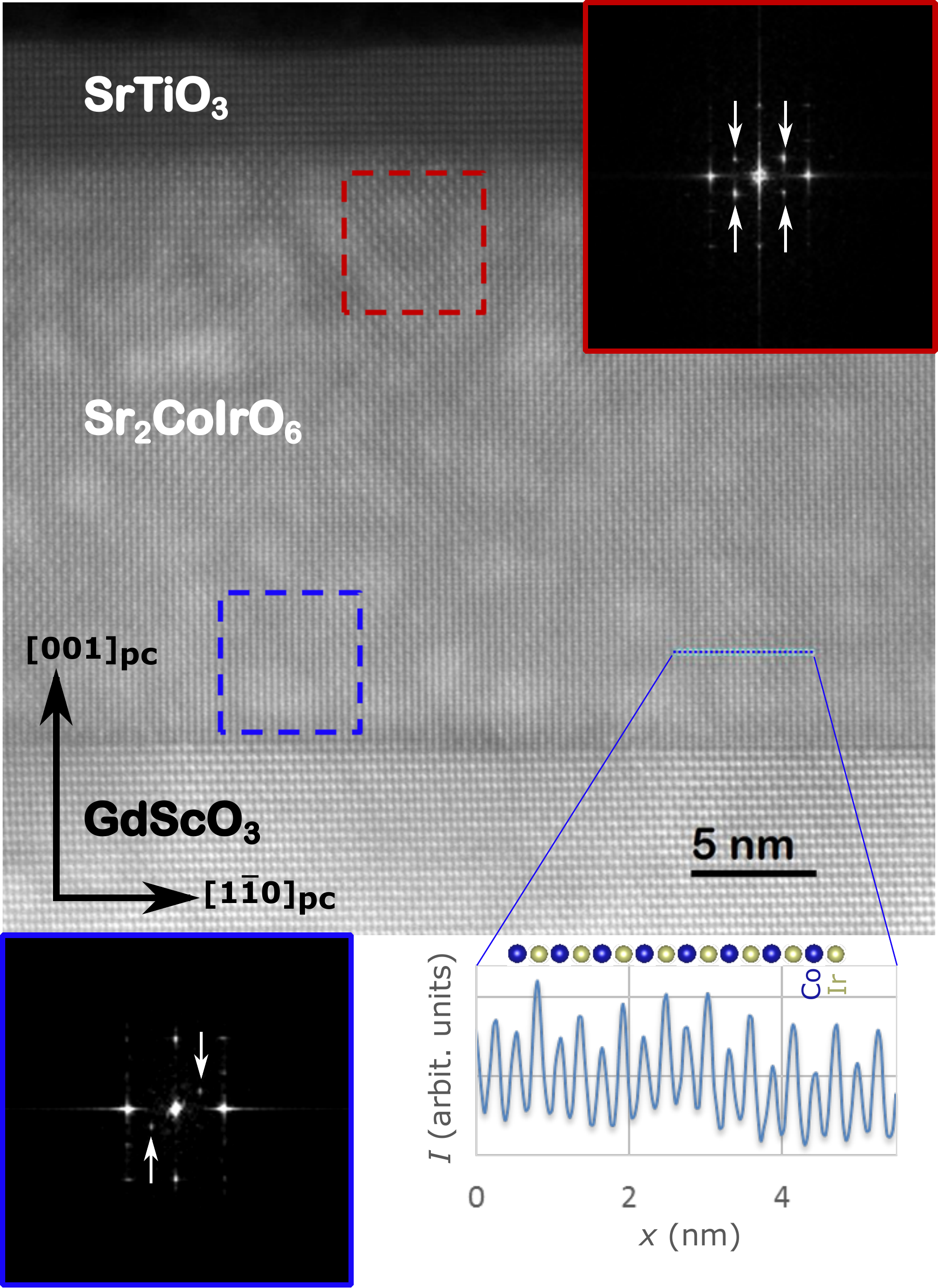}
\caption{\label{Fig_TEM_110_GSO}HAADF STEM image in [110]$_\text{pc}$ direction of a SCIO thin film on (001)$_\text{c}$ oriented GSO including a line scan proving the alternating Co/Ir arrangement. Local FFTs (red and blue squares) indicate areas with good (red) and worse (blue) B-site ordering. The spots marked by white arrows are of prime importance for the B-site ordering.}
\end{figure}

To verify the B-site ordering of the samples on the micro scale, HAADF STEM measurements in [110]$_\text{pc}$ direction were performed on the SCIO samples grown on (001)$_\text{pc}$ oriented NGO, STO and GSO (see Figure \ref{Fig_TEM_110_NGO}, \ref{Fig_TEM_110_STO} and \ref{Fig_TEM_110_GSO}). As shown in Figure \ref{Fig_Crystal_110} this special direction provides an alternating arrangement of Co and Ir pillars only in a well ordered double perowskite structure.

Due to the good $Z$-contrast \cite{S_Hartel} ($I\propto Z^{1.6-1.9}$) of the HAADF mode, the difference between Co and Ir is clearly visible. The theoretical intensity ratio is $I_\text{Co} : I_\text{Ir} \approx 1 : 8$ and therefore line scans can already provide the desired information about the B-site ordering. 

In our case most parts of the sample seems to be in a well ordered state, supporting the results shown in the main text. 

In order to estimate the degree of B-site ordering we have performed a simulation of HR-STEM images for different occupancies of the corresponding columns with Co and Ir atoms. The results are shown in Fig. S3. The simulations allow us to estimate the degree of Co/Ir ordering to be 80-90\%. Moreover, our HR-STEM images also show areas which do not have pronounced difference in the intensity of Ir/Co atomic columns. The corresponding FFT pattern also do not reveal the superstructure spots. In general, TEM is an essentially local method. However, the images of all studied films demonstrate a very similar ratio of ordered and disordered areas, which is at least $4:1$. Thus, we can conclude that the degree of Co/Ir ordering is higher than 65\%. Similar results were also reported for the bulk SCIO material (13.2\% foreign B species occupation)~\cite{S_Narayanan}.

\newpage

\subsection*{Identification of foreign phase within the SCIO sample for SQUID measurements and estimation of its relative volume fraction}
\begin{figure}[h]
\includegraphics[width=0.90\linewidth]{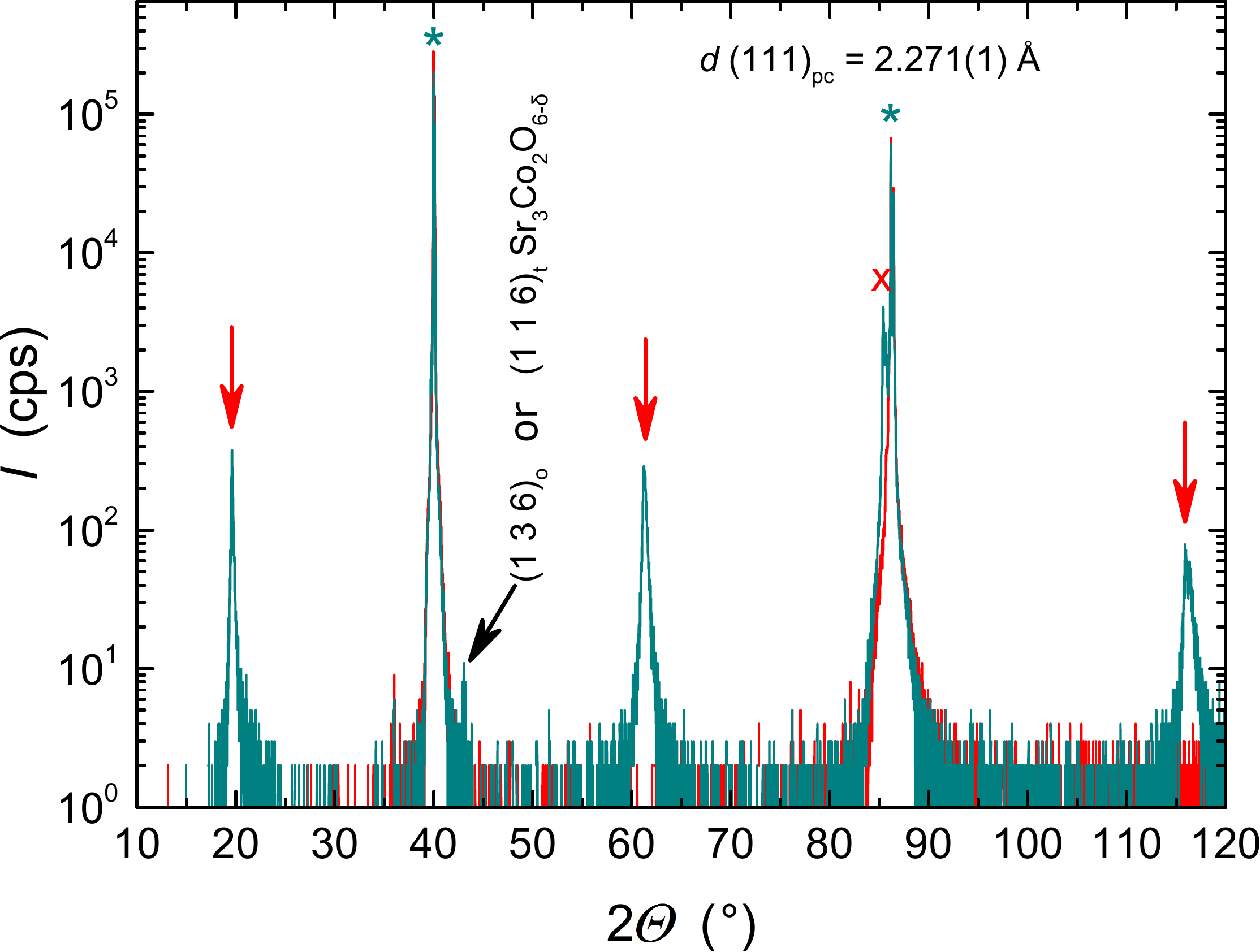}
\caption{\label{Fig_XRD_foreign_phase}$\theta-2\theta$ XRD scan: Green stars indicate peaks from the (111)-STO substrate, the red cross shows the (222)-SCIO reflex and the red arrows mark peaks due to the ordered superstructure. The black arrow points to the peak from the foreign phase, which is identified as Sr$_3$Co$_2$O$_{6-\delta}$.}
\end{figure}

Figure \ref{Fig_XRD_foreign_phase} shows the $\theta-2\theta$ XRD pattern of the $\sim 250\,\text{nm}$ thick (111)$_\text{pc}$ oriented SCIO sample, which was used to probe the magnetic properties with the SQUID magnetometer. The additional peak at $2\theta = 43.0(3)\,^\circ$ could be identified as tetragonal or orthorhombic Sr$_3$Co$_2$O$_{6-\delta}$ (SCO). The further non substrate peaks belong to (111)$_\text{pc}$ oriented Sr$_2$CoIrO$_6$.
The clearly visible super lattice peaks (marked by red arrows) indicate well developed B-site ordering. The thin film is still fully strained, as evidenced by the extracted out-of-plane lattice constant, which is the same as for thinner samples.

\begin{figure}[t]
\includegraphics[width=0.95\linewidth]{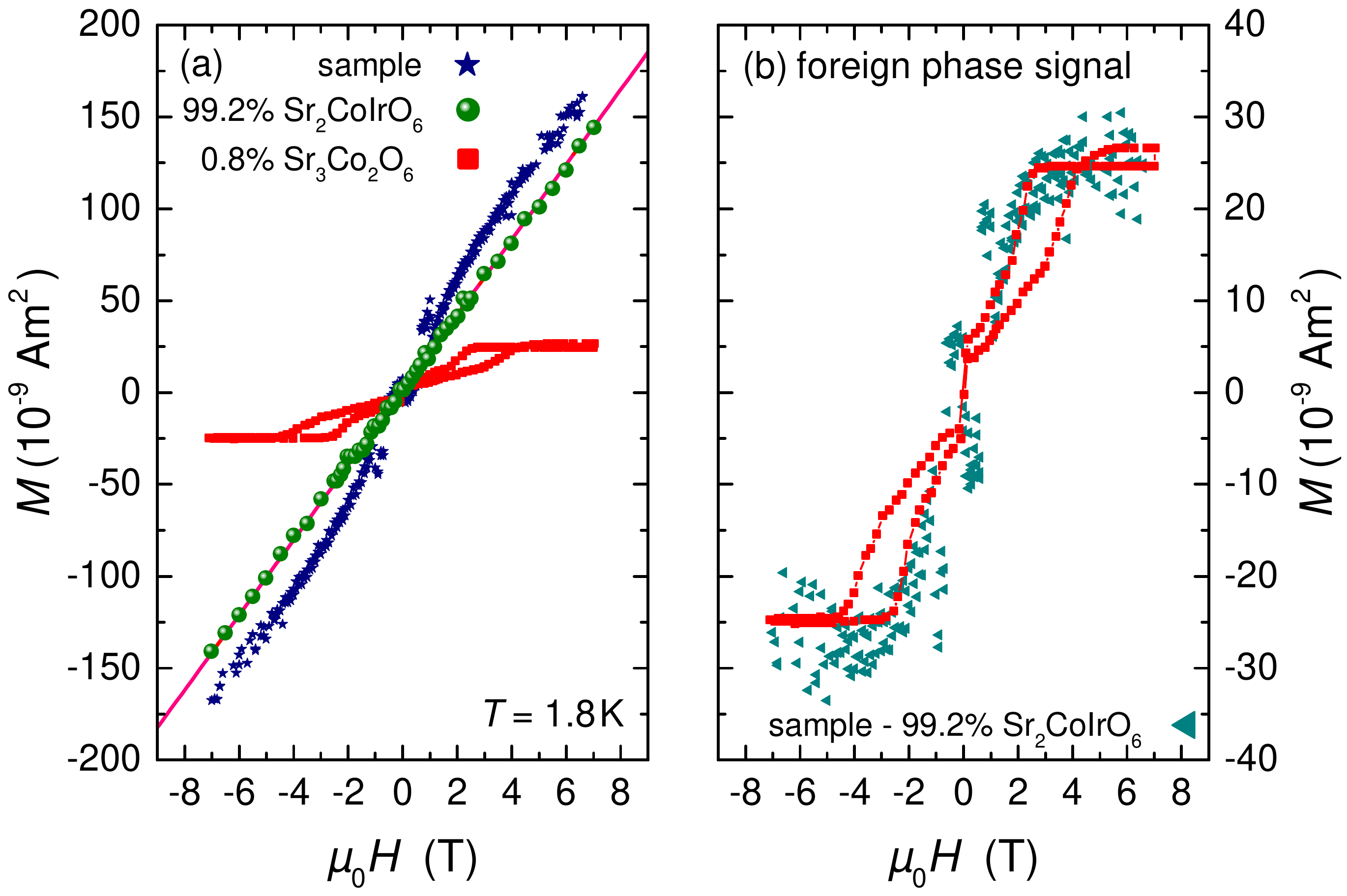}
\caption{\label{Fig_MPMS_MvsH} (a) Field dependence of the magnetization of a $\sim 250\,\text{nm}$ thick (111)$_\text{pc}$ oriented sample (blue stars) at lowest $T$ with the field applied in the film direction, including the volume weighted signals of SCIO fraction of the sample (green balls), literature data of SCIO for the same volume fraction (red solid line) from reference \cite{S_Narayanan} and literature data of Sr$_3$Co$_2$O$_{6-\delta}$ from reference \cite{S_Wang} for the Sr$_3$Co$_2$O$_{6-\delta}$ fraction. (b) Signal of the Sr$_3$Co$_2$O$_{6-\delta}$ fraction of the sample (cyan triangles) in comparison to literature data for the same volume fraction (red squares).}
\end{figure}

For a quantitative determination of the volume fraction of the Sr$_3$Co$_2$O$_{6-\delta}$ foreign phase we could use the literature data  of the field dependence of its magnetization from reference \cite{S_Wang}. To describe our measured thin film magnetization (blue stars in Figure \ref{Fig_MPMS_MvsH} (a))  a volume weighted superposition of SCIO and Sr$_3$Co$_2$O$_{6-\delta}$ has been assumed. The estimated foreign phase fraction is $0.8\,\%$ (see Figure \ref{Fig_MPMS_MvsH} (b)) and the pure SCIO signal follows in good agreement with the literature data from reference \cite{S_Narayanan} a linear behavior.

\begin{figure}[b]
\includegraphics[width=0.95\linewidth]{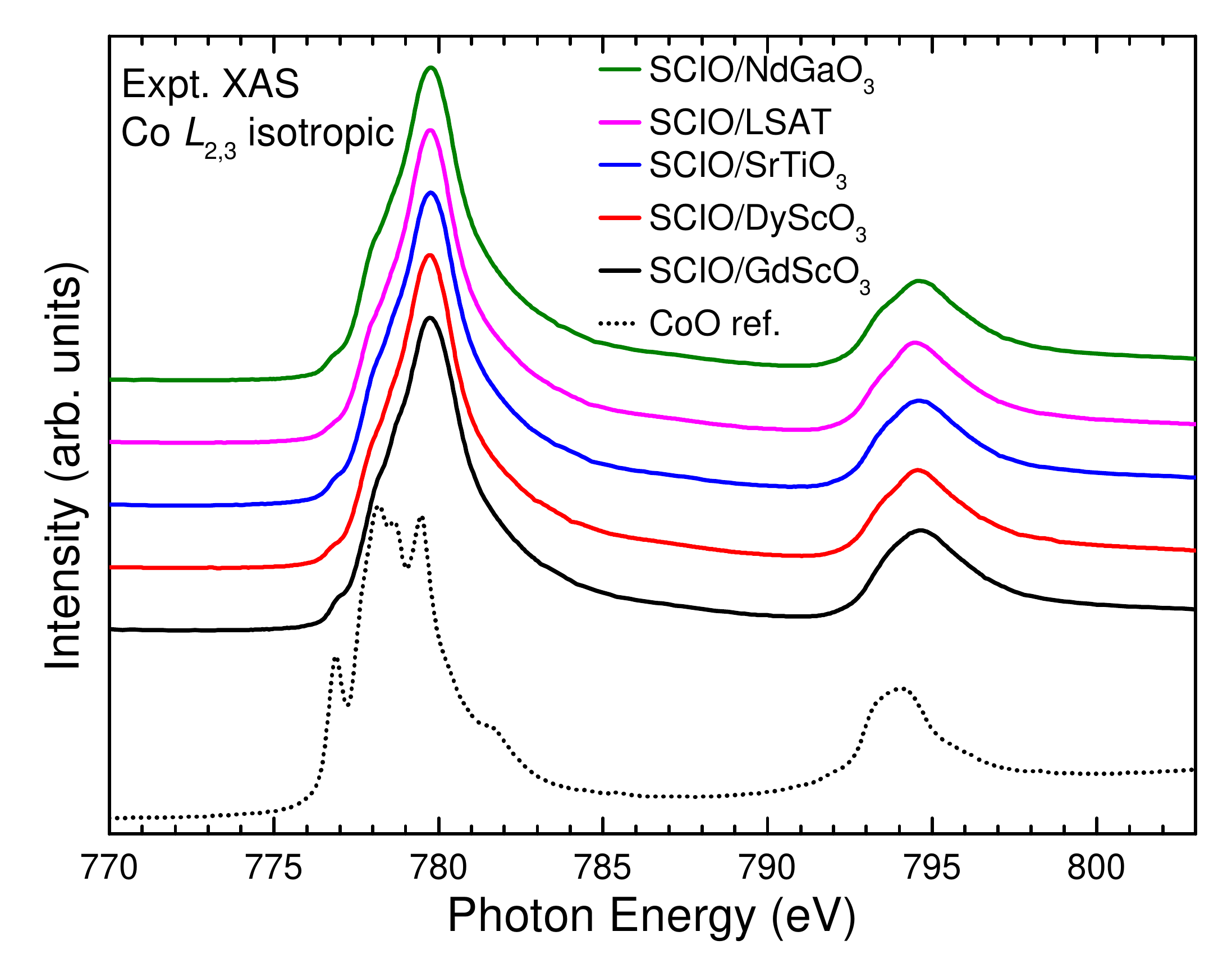}
\caption{\label{Fig_S_XAS_iso}Experimental isotropic XAS spectra without subtracting Co$^{2+}$ impurities.}
\end{figure}

\subsection*{XAS spectra and configuration interaction cluster calculation}
Figure~\ref{Fig_S_XAS_iso} shows the experimental isotropic Co $L_{2,3}$ XAS spectra of the SCIO thin films on GSO, DSO, STO, LSAT and NGO substrates taken at 300 K,
\begin{figure*}
\includegraphics[width=0.95\linewidth]{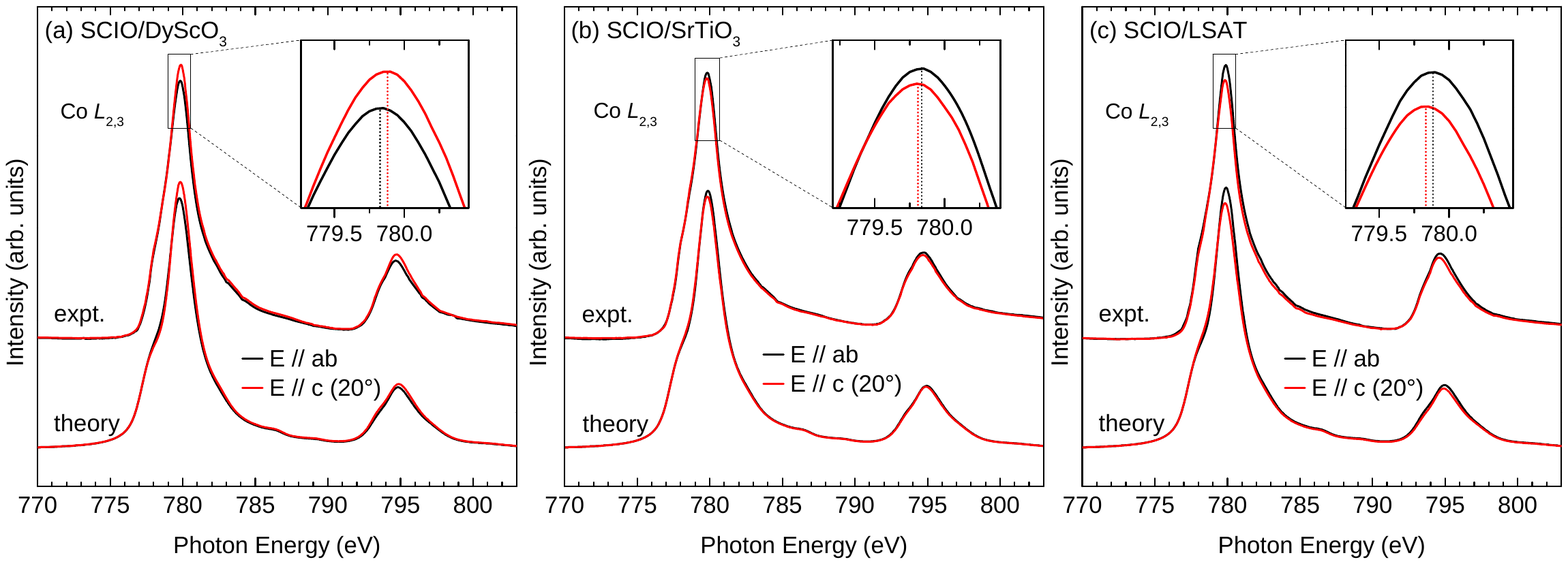}
\caption{\label{Fig_S_XAS_LD}Experimental and theoretical polarization-dependent XAS spectra of the SCIO thin films on DSO, STO, and LSAT substrates.}
\end{figure*}
together with a standard Co$^{2+}$ spectrum from CoO. The foot of the $L_3$ main line of the SCIO thin films at 776.9 eV is due to Co$^{2+}$ contributions. Comparing the intensity ratio of the foot (776.9 eV) and the $L_3$ main line (779.74 eV), we derive 10~\%, 7~\%, 9~\%, 6~\%, and 8~\% of Co$^{2+}$ impurities from the SCIO thin films on GSO, DSO, STO, LSAT and NGO substrates, respectively. The isotropic XAS spectra were obtained via the formula of $I = I_{\parallel} + 2I_{\perp}$, where $I_{\parallel}$ is the spectrum with the E parallel to the [001]$_{pc}$ surface normal extracted from $I_{\parallel} = [I_{\parallel c (20^\circ)} - I_{\perp}\cos^2(70^{\circ})]/\sin^2(70^{\circ})$. $I_{\parallel c (20^\circ)}$ and $I_{\perp}$ are the spectra measured with E $\parallel$ 20$^{\circ}$ off the [001]$_{pc}$ surface normal and E $\perp$ the [001]$_\text{pc}$ surface normal, i.e. E $\parallel$ ab, respectively.

The x-ray absorption (XAS) spectra were calculated using the well established configuration interaction cluster model that includes the full atomic multiplet theory and the local effects of the solid \cite{S_deGroot,S_Tanaka}. It accounts for the intra-atomic Co $3d-3d$ and $2p-3d$ Coulomb and exchange interactions, the atomic $2p$ and $3d$ spin-orbit couplings, the O $2p$-Co $3d$ hybridization and the local ionic crystal field. The calculations were done using the program XTLS 8.3 \cite{S_Tanaka}. We have considered a CoO$_6$ cluster. Parameters for the multipole part of the Coulomb interactions were given by 80\% of the Hartree-Fock values for the $d-d$ and $p-d$ Slater integrals, while the monopole parts ($U_{dd}$, $U_{cd}$) as well as the O $2p$-Co $3d$ charge transfer energy were adopted from typical values for Co$^{3+}$ \cite{S_PRL92,S_PRL102,S_PRB95}. The hopping integrals between the Co $3d$ and O $2p$ were calculated for the various Co-O bond lengths according to Harrison's description. The ionic crystal field splitting of the $t_{2g}$ and $e_g$ levels is set to be 0.5 eV for all SCIO thin films. Due to the epitaxial strain from the substrates, the $e_g$ levels are split into a $d_{3z^{2}-r^{2}}$ and a $d_{x^{2}-y^{2}}$ with an energy difference of $D_{eg}$, and the $t_{2g}$ levels are split into a doublet ($d_{xz}$ and $d_{yz}$) and a singlet ($d_{xy}$) with an energy difference of $D_{t2g}$. The $D_{eg}$ and $D_{t2g}$ were tuned to fit the experimental spectra: $D_{eg}$ = -0.25 (GSO), -0.2 (DSO), 0.1 (STO), 0.2 (LSAT), and 0.25 (NGO) eV; $D_{t2g}$ = -0.03 (GSO), -0.018 (DSO), 0.008 (STO), 0.019 (LSAT), and 0.023 (NGO) eV. We show the experimental and calculated polarization-dependent Co $L_{2,3}$ XAS spectra of the SCIO thin films on DSO, STO, and LSAT substrates in Figure~\ref{Fig_S_XAS_LD} (a), (b), and (c), respectively. Those results for the thin films on GSO and NGO are shown in Figure~7 of the main text. We argue that these XAS derived parameters can be used for the analysis of the low temperature magnetic properties since the change of the anisotropy in the crystal field parameters in going from room temperature to low temperatures due to thermal contraction of the substrates is negligible as we will explain now. From literature we can find the following thermal expansion coefficients for the substrates: GSO=1.09x10$^{-5}$ (K$^{-1}$), DSO=0.84x10$^{-5}$ (K$^{-1}$), STO=0.9x10$^{-5}$ (K$^{-1}$), LSAT=0.82x10$^{-5}$ (K$^{-1}$), NGO=0.9x10$^{-5}$ (K$^{-1}$). We then can estimate the substrate lattice constants at, for example, 5~K. Knowing that the SCIO films are fully strained, and knowing their Poisson ratio (from the data displayed in Fig. 4) together with the volume change of the bulk SCIO from 300~K (238.56 $\mathring{\text{A}}^3$) to 5~K (237.09 $\mathring{\text{A}}^3$) \cite{S_Mikhailova}, we can subsequently estimate the out-of-plane lattice constants of the SCIO films at 5~K. We thus find that the c/a ratio at 5~K on GSO, DSO, STO, LAST, and NGO is 0.9766, 0.9855, 1.0141, 1.0363, and 1.045, respectively. This is marginally different from the values at 300~K, namely 0.9751, 0.9858, 1.0141, 1.0369, and 1.0451. Thus the anisotropy due to the crystal field at 5~K is practically the same as that at 300K. Please be noted that for comparing with the calculated spectra the experimental spectra in Figure~\ref{Fig_S_XAS_LD} (a), (b), and (c) have been subtracted by 7~\%, 9~\% and 6~\% of Co$^{2+}$ impurities, respectively. For the simulations of the magnetic anisotropy energy as a function of the spin direction, an exchange field is applied on the Co ion.
\vspace*{-1em}
\subsection*{Zero field transport properties}

\begin{figure}[t]
\includegraphics[width=0.95\linewidth]{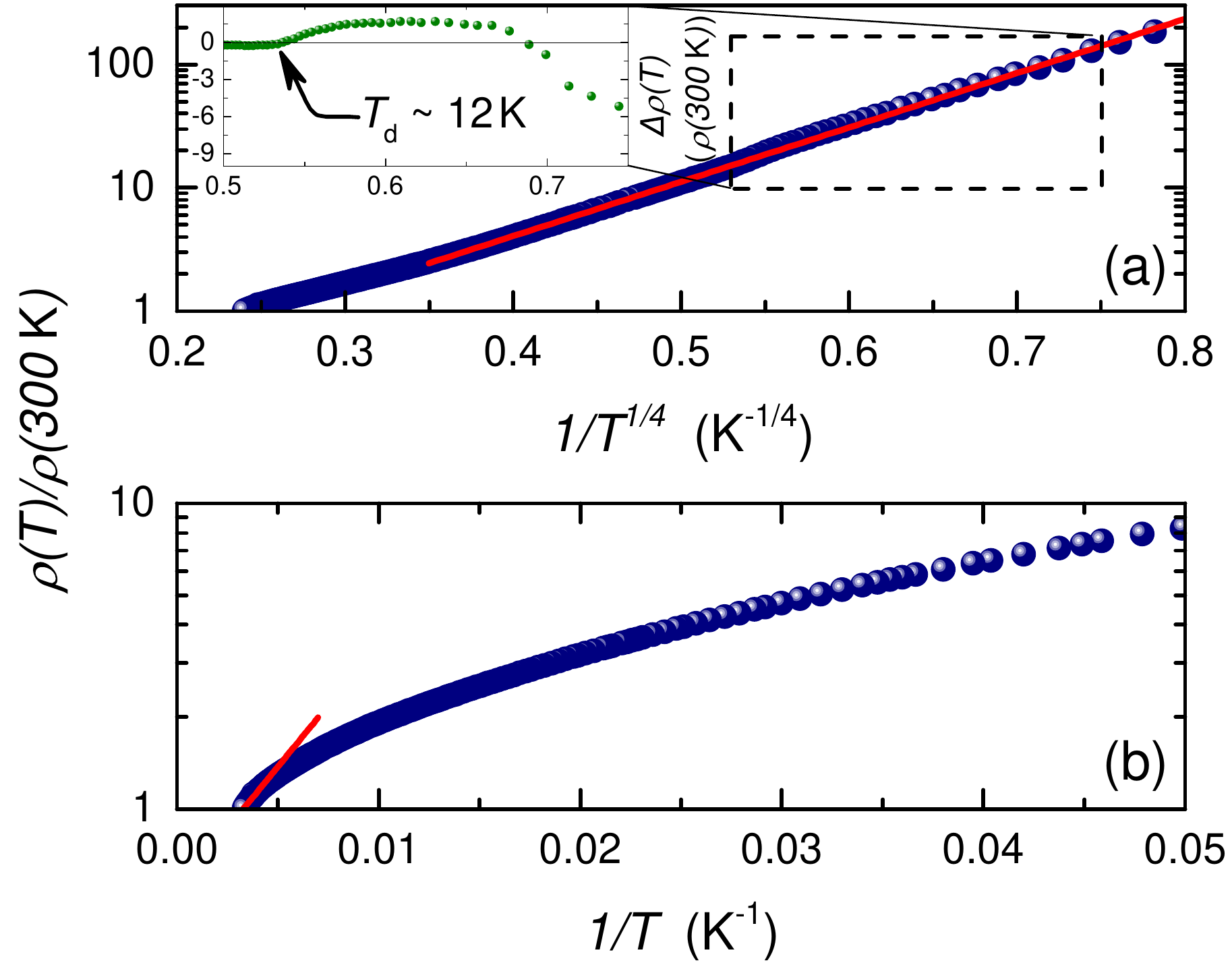}
\caption{\label{Fig_RT-111-Supp}Electrical resistivity ratio of a (111)$_\text{pc}$ oriented sample on STO with STO protection on log-scale as function of $T^{-1/4}$. The red line indicates three-dimensional variable range hopping (VRH) behavior. The inset displays the deviation of the data from a VRH fit. The arrow at $T_d=12$~K marks the temperature, below which deviation from VRH behavior occurs. (b) Arrhenius plot for the same data. The red line indicates thermally activated electrical resistivity behavior between 250 and 300~K with a gap of $32(5)$~meV.}
\end{figure}

\begin{figure}[t]
\includegraphics[width=0.95\linewidth]{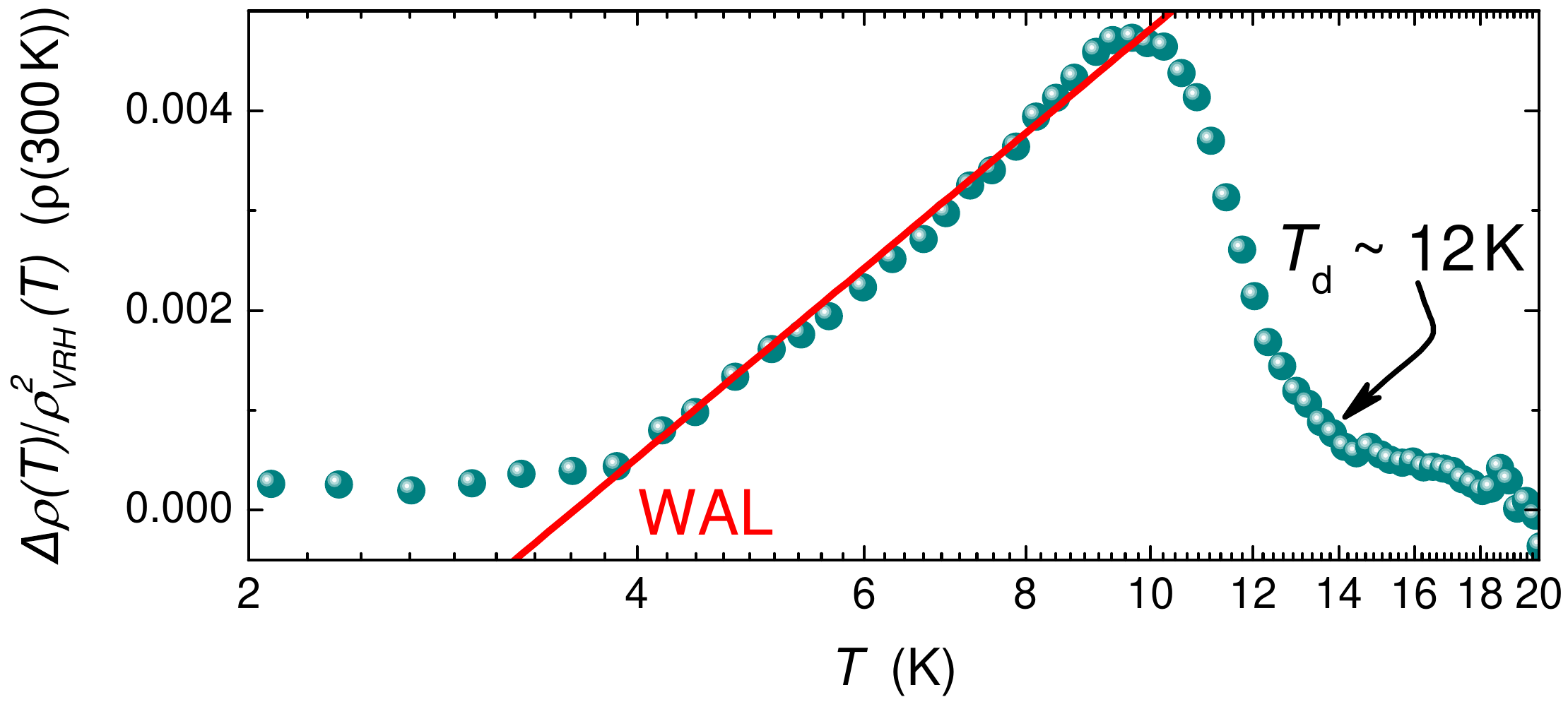}
\caption{\label{Fig_RT-111-Supp-2} Temperature dependent deviation of the electrical resistivity from VRH behavior in Figure \ref{Fig_RT-111-Supp} (a) (as $\rho_\text{VRH}$) as $\Delta\rho/\rho_\text{VRH}^2$ vs $\log T$.}
\end{figure}

We analysed the high temperature electrical resistance behavior of the air protected thin film in a simple thermal activation scenario as shown in Figure \ref{Fig_RT-111-Supp} (b) and extracted as transport gap size $\Delta E_\text{trans.} = 32(5)\,\text{meV}$. The deviation from Arrhenius type behavior sets in at $T \leq 250\,\text{K}$ and for $T \leq 70\,\text{K}$ we could describe our measurement in a good agreement with a three-dimensional variable-range hopping (VRH) model as introduced by Mott for disordered systems \cite{S_Mott} (see Figure \ref{Fig_RT-111-Supp} (a)). At $T_\text{d} \sim 12\,\text{K}$ we observe a small kink and deviation from VRH (see inset of Figure \ref{Fig_RT-111-Supp} (a) and Figure \ref{Fig_RT-111-Supp-2}). This deviation is not induced by the quantum correction due to weak anti-localization (WAL), because such correction would lead to a negative contribution (positive derivative of $\Delta\rho / \rho_\text{VRH}^2$) being logarithmic in temperatu
 re \cite{S_Bergmann,S_Anderson}, as found below $T=10\,\text{K}$ (see  Figure \ref{Fig_RT-111-Supp-2}). Thus, it is likely, that the positive contribution results from the magnetic ordering of the Co sublattice, as it could be found in a similar way  in the resistivity along the c-axis $\rho_\text{c}$ in the related perovskite Sr$_2$IrO$_4$ \cite{S_Fruchter}, but this assumption needs to be substantiated by further investigations.

\begin{figure}[t]
\includegraphics[width=0.95\linewidth]{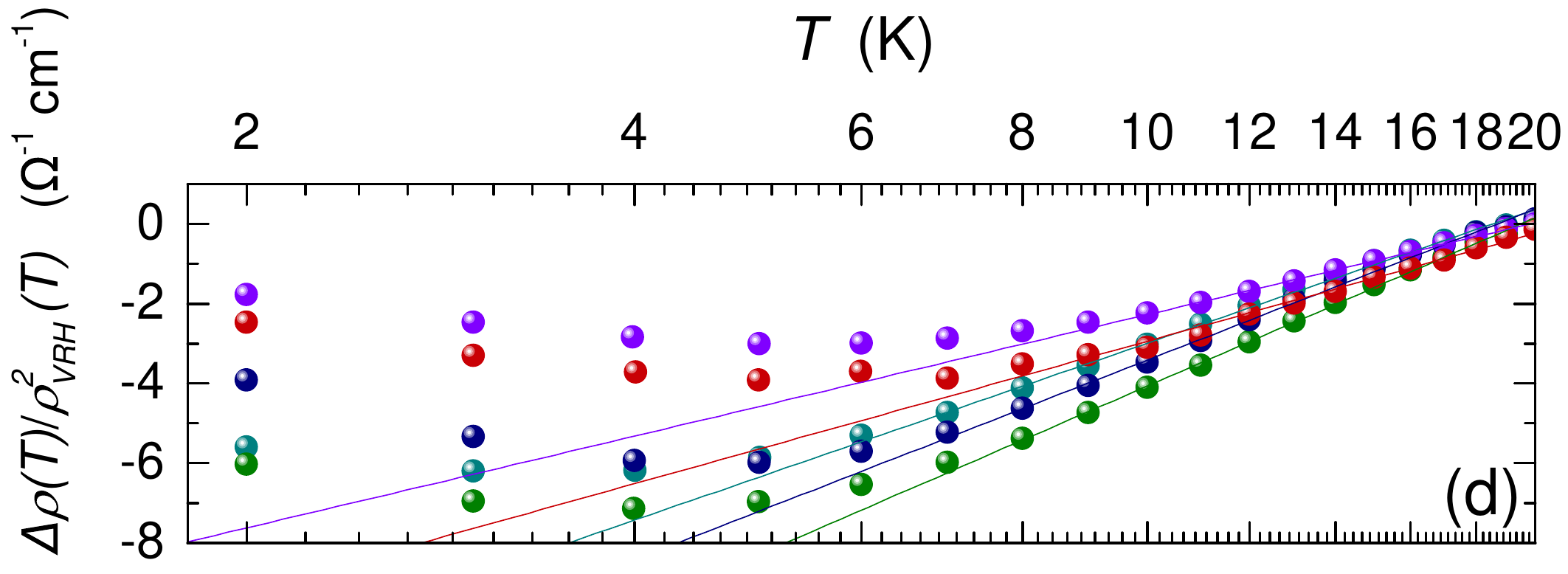}
\includegraphics[width=0.95\linewidth]{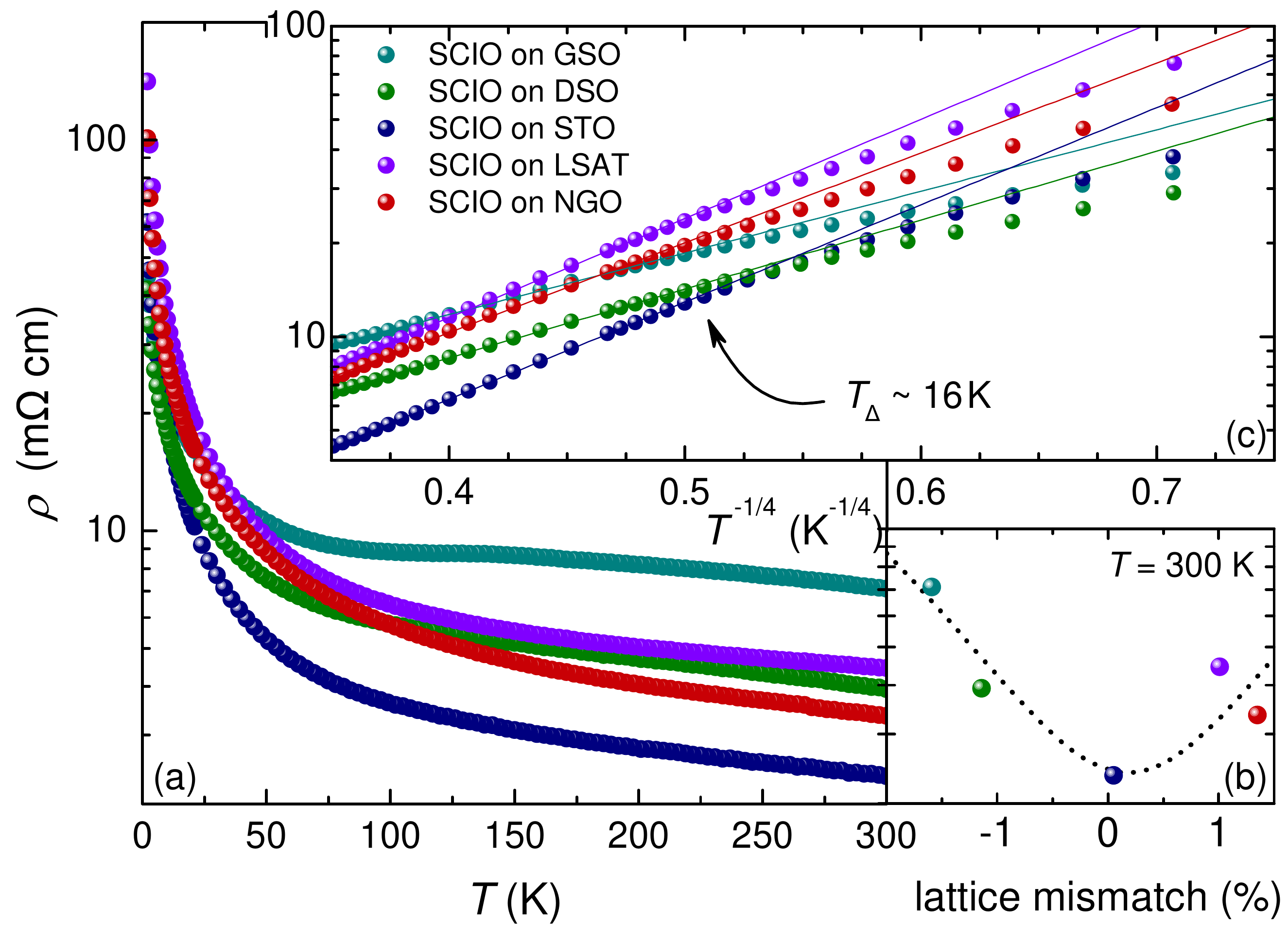}
\caption{\label{Fig_RT-001}(a) Temperature dependence of all samples of the (001)$_\text{pc}$ strain series and representation for 3D VRH-scenario (c) and its low temperature deviation (d). (b) Strain dependence of the $\rho(300\,\text{K})$ value.}
\end{figure}

Figure \ref{Fig_RT-001} (a) shows  temperature dependent transport measurements of various (001)$_\text{pc}$ strained samples. The overall insulating behavior is not changed by epitaxial strain and also the deviation temperature $T_\Delta \sim 16\,\text{K}$ from the low temperature 3D VRH scenario does not change significantly (see Figure \ref{Fig_RT-001} (b)). Our analysis of the high temperature behavior within the simple thermal activation of electronic transport scenario reveals a transport gap size $\Delta E_\text{trans.} \sim 10(2)\,\text{meV}$ which is reduced compared to the (111)$_\text{pc}$ oriented sample.

Another obvious effect of the strain is  the  evolution of the specific resistance at room temperature $\rho(300\,\text{K})$, which increases with increasing absolute strain strength $\sim f^2$, whereby $f$ is the lattice mismatch (see Figure \ref{Fig_RT-001} (c)). Compared to other transition metal oxides this effect is not unexpected and was found (with other power law behavior) also in La$_{0.7}$Ca$_{0.3}$MnO$_3$ \cite{S_Wu} and ZnO \cite{S_Ghosh}.

\subsection*{Magnetotransport for (001)$_\text{pc}$ strained SCIO}
\begin{figure}[t]
\includegraphics[width=0.95\linewidth]{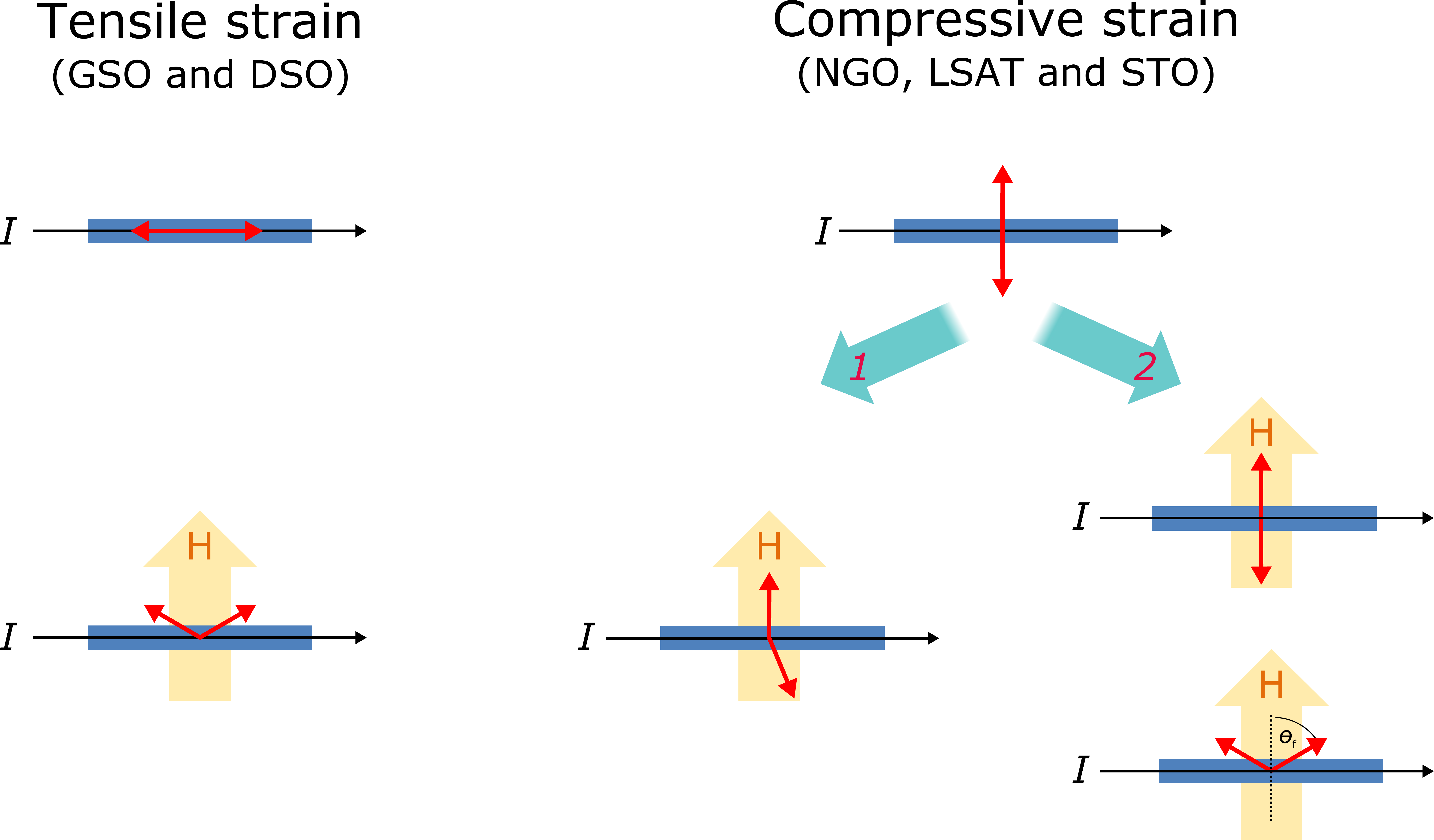}
\caption{\label{Fig_MR-modell}Illustration of the utilized models for the magnetic response of the AFM ordered Co sublattice in case of an external field perpendicular to the thin film plane. The black and red arrows indicate the current direction and moment orientation, respectively. For compressive strain, two different scenarios indicated by the wide blue arrows are considered. See text.}
\end{figure}

Historically motivated by the Dresselhaus effect the WAL quantum corrections leads to \cite{S_Dresselhaus}
\begin{align}
	\Delta\sigma(H) = \frac{e^2}{2\pi^2\hbar}\left[\frac{3}{2}\cdot\mathcal{F}\left(\frac{H}{4H_\text{so}+H_\phi}\right)-\frac{1}{2}\cdot\mathcal{F}\left(\frac{H}{H_\phi}\right)\right]
\end{align}
in which we assume isotropic spin orbit scattering and where $\mathcal{F}(x) = \ln(x)+\Psi(\frac{1}{2}+\frac{1}{x})$ with the Digamma function $\Psi$. Consequently the WAL will influence the magnetoresistance $MR$ as follows
\begin{align}
MR_\text{WAL} = \frac{R(H)-R(0)}{R(0)} = \frac{1}{1+\rho(0)/d\cdot\Delta\sigma(H)}-1 \label{eq:WAL:pure}
\end{align}
Furthermore at lowest temperature the AFM ordering of the Co sublattice and its response to external magnetic field by the anisotropic magnetoresistance (AMR) effect needs to be taken into account. For the description of the field dependence of the AMR effect we follow the Ansatz of Birss \cite{S_Birss} and Muduli \textit{et al.} \cite{S_Muduli}. The components $\rho_{ij}$ of the resistance tensor are written in a taylor expansion of the components of the magnetization direction $m_i^{(\pm)}$ for each of the two Co sublattices ($\pm$)
\begin{align}
	\rho_{ij} = a_{ij}+\sum_\pm\left[\sum_{k=1}^3{a_{ijk}m_k^{(\pm)}}+\sum_{(l,k)=1}^3{a_{ijkl}m_k^{(\pm)}m_l^{(\pm)}}+\ldots\right]\label{eq:amr:taylor}
\end{align}
in which the expansion coefficients $a_{ij\ldots}$ have to preserve the lattice symmetry \cite{S_Limmer}. Due to the Onsager theorem equation (\ref{eq:amr:taylor}) could be divided into a symmetric and an antisymmetric part, where the symmetric part describes the generalized magnetoresistance and the antisymmetric part the generalized Hall effect \cite{S_McGuire}. For in-plane current direction, a $D_{4h}$ symmetry and Taylor expansion until second order in magnetization, the symmetric part of equation (\ref{eq:amr:taylor}) reduces to
\begin{align}
	\rho^\text{s} \approx a_{11}+\sum_\pm\vec{m}^{(\pm)}\cdot\mathcal{A}\cdot\vec{m}^{(\pm)} \label{eq:amr:symmetric}
\end{align}
where $\mathcal{A}$ is a diagonal matrix. With the assumption that the field dependence of equation (\ref{eq:amr:symmetric}) is mainly located in the magnetization orientation, the AMR effect will influence the magnetoresistance $MR$ as follows
\begin{align}
MR_\text{AMR} \approx \frac{1}{\rho(0)}\sum_\pm\big[&\vec{m}^{(\pm)}(H)\cdot\mathcal{A}\cdot\vec{m}^{(\pm)}(H)\nonumber\\
-&\vec{m}^{(\pm)}(0)\cdot\mathcal{A}\cdot\vec{m}^{(\pm)}(0)\big]\label{eq:amr:final}
\end{align}
which leads in total to
\begin{align}
	MR = MR_\text{WAL}+MR_\text{AMR}
\end{align}
We now need to model the magnetic response of the AFM ordered Co sublattice in case of an external field perpendicular to the thin film plane. Due to the change of magnetic easy axis with inversion of the strain direction we need to setup a separate model for each strain direction. Furthermore we have to distinguish between to scenarios in case of the external magnetic field parallel to the magnetic easy axis (quasi independent FM sublattices vs. classical spin-flip). Figure \ref{Fig_MR-modell} illustrates all of our scenarios.

For the tensile strain case we assume, that the angle between the magnetization and the current direction is proportional to the external magnetic field and reaches $90\,^\circ$ at the saturation field $H_\text{s}$. Then the magnetisation direction could be described by
\begin{align}
\vec{m}^{(\pm)}_\text{tensile}(H) &=\nonumber\\ \vec{e}_\text{oop}&\cdot\left[\sin\left(\frac{\pi}{2}\cdot\frac{|H|}{H_\text{s}}\right)\cdot\Theta\left(H_\text{s}-|H|\right)+\Theta\left(|H|-H_\text{s}\right)\right]\nonumber\\
\pm\vec{e}_\text{ip}&\cdot\cos\left(\frac{\pi}{2}\cdot\frac{|H|}{H_\text{s}}\right)\cdot\Theta\left(H_\text{s}-|H|\right)
\end{align}
where $\vec{e}_\text{oop}$ ($\vec{e}_\text{ip}$) is a unit vector in out-of-plane (in-plane) direction and $\Theta$ the Heaviside step function. The field dependence of the MR in tensile strained case therefore leads to
\begin{align}
	MR \approx ~&\frac{1}{1+\frac{\rho(0)}{d}\cdot\Delta\sigma(H)}-1\nonumber\\+~&a^\text{ip}\cdot\left[\sin^2\left(\frac{\pi}{2}\cdot\frac{|H|}{H_\text{s}}\right)\cdot\Theta(H_\text{s}-|H|)+\Theta(|H|-H_\text{s})\right]\label{eq:MR:C-typ}
\end{align}
where $a^\text{ip}  = \frac{2\mathcal{A}_{33}-(\mathcal{A}_{11}+\mathcal{A}_{22})}{\rho(0)}$.

\begin{figure}[t]
\includegraphics[width=\linewidth]{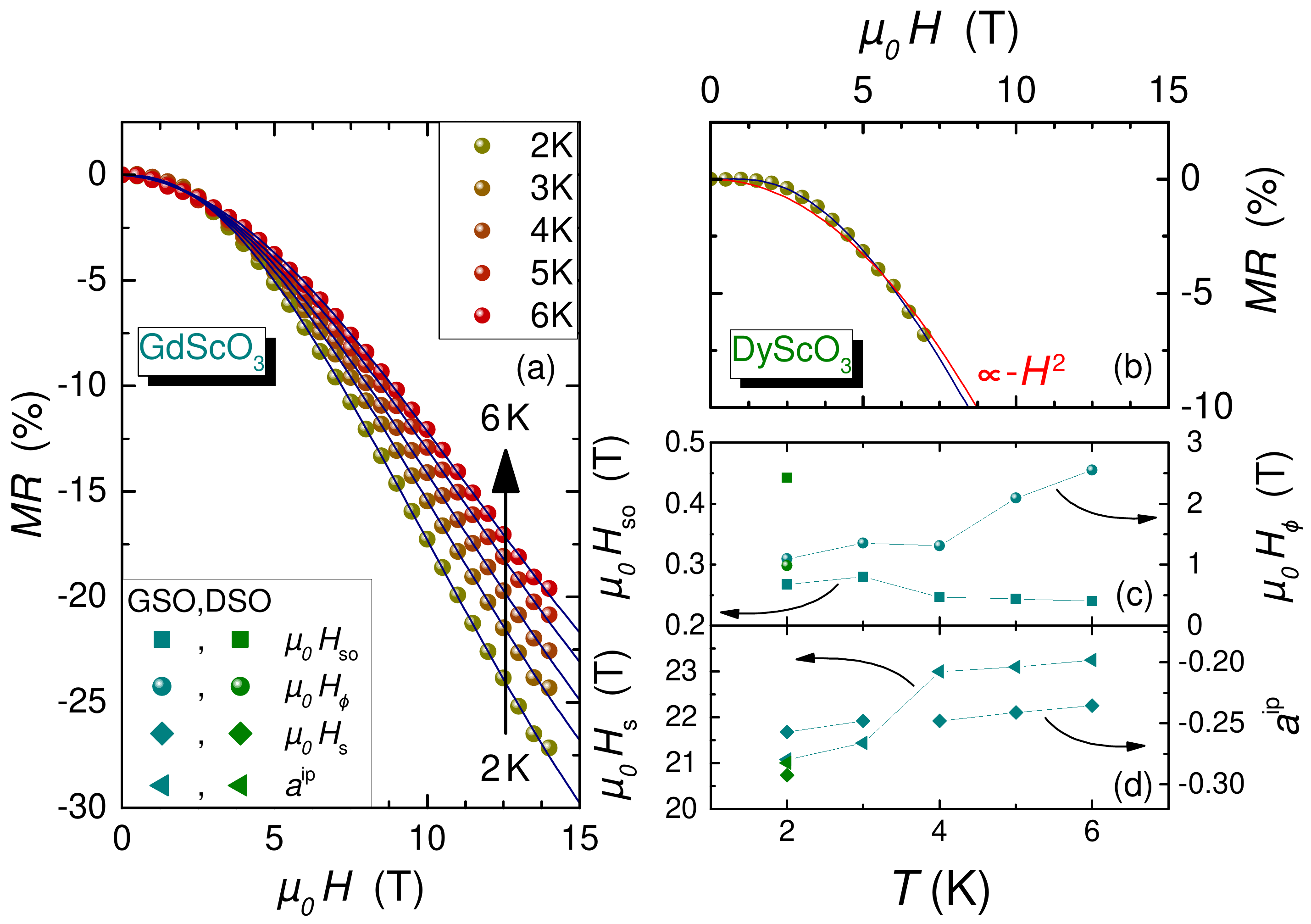}
\caption{\label{Fig_MR-tensile}Field dependence of the MR of tensile strained SCIO thin films on GSO (a) and DSO (b) including fits according to equation (\ref{eq:MR:C-typ}). (c) and (d) Temperature dependence of the fit parameters.}
\end{figure}

This model enables us to describe in good consistency the collected MR data which are shown in Figure \ref{Fig_MR-tensile}.
As expected the spin-orbit scattering field $H_\text{so}$ is nearly temperature independent, whereas the inelastic scattering field $H_\phi \sim T^p$ follows a power law behavior with $p\approx1$. The saturation field $H_\text{s}$ also slightly increases with temperature. This is reasonable because in the mean field approximation the argument of the Brillouin function $y\sim H/T$ needs to be constant to reach the saturation.

In case of compressive epitaxial strain we have to distinguish between two possibilities. In the first scenario the AFM coupling between the two sublattices is very weak, which enables us to deal with two quasi independent FM sublattices. In this scenario only the ($-$) sublattice, where the magnetization is aligned antiparallel to the external field, is influenced and has to be taken into account in equation (\ref{eq:amr:final}). As in the tensile strained case we model its magnetization direction by a linear coupling of the external magnetic field to the angle between magnetization and the current direction. This leads to
\begin{align}
	\vec{m}^{(-)}_\text{comp. I }(H) &=\nonumber\\ \vec{e}_\text{oop}&\cdot\left[\Theta\left(|H|-H_\text{s}\right)-\cos\left(\pi\cdot\frac{|H|}{H_\text{s}}\right)\cdot\Theta\left(H_\text{s}-|H|\right)\right]\nonumber\\
+\vec{e}_\text{ip}&\cdot\sin\left(\pi\cdot\frac{|H|}{H_\text{s}}\right)\cdot\Theta\left(H_\text{s}-|H|\right)
\end{align}
and thereby to the following field dependence of MR
\begin{align}
	MR\approx &\frac{1}{1+\frac{\rho(0)}{d}\cdot\Delta\sigma(H)}-1\nonumber\\+&a^\text{oop}\cdot\sin^2\left(\pi\cdot\frac{|H|}{H_\text{s}}\right)\cdot\Theta(H_\text{s}-|H|)\label{eq:MR:A-typ:1}
\end{align}
where $a^\text{oop}  = \frac{\mathcal{A}_{11}+\mathcal{A}_{22}-2\mathcal{A}_{33}}{2\rho(0)} = -\frac{a^\text{ip}}{2}$.
\begin{figure}[t]
\includegraphics[width=\linewidth]{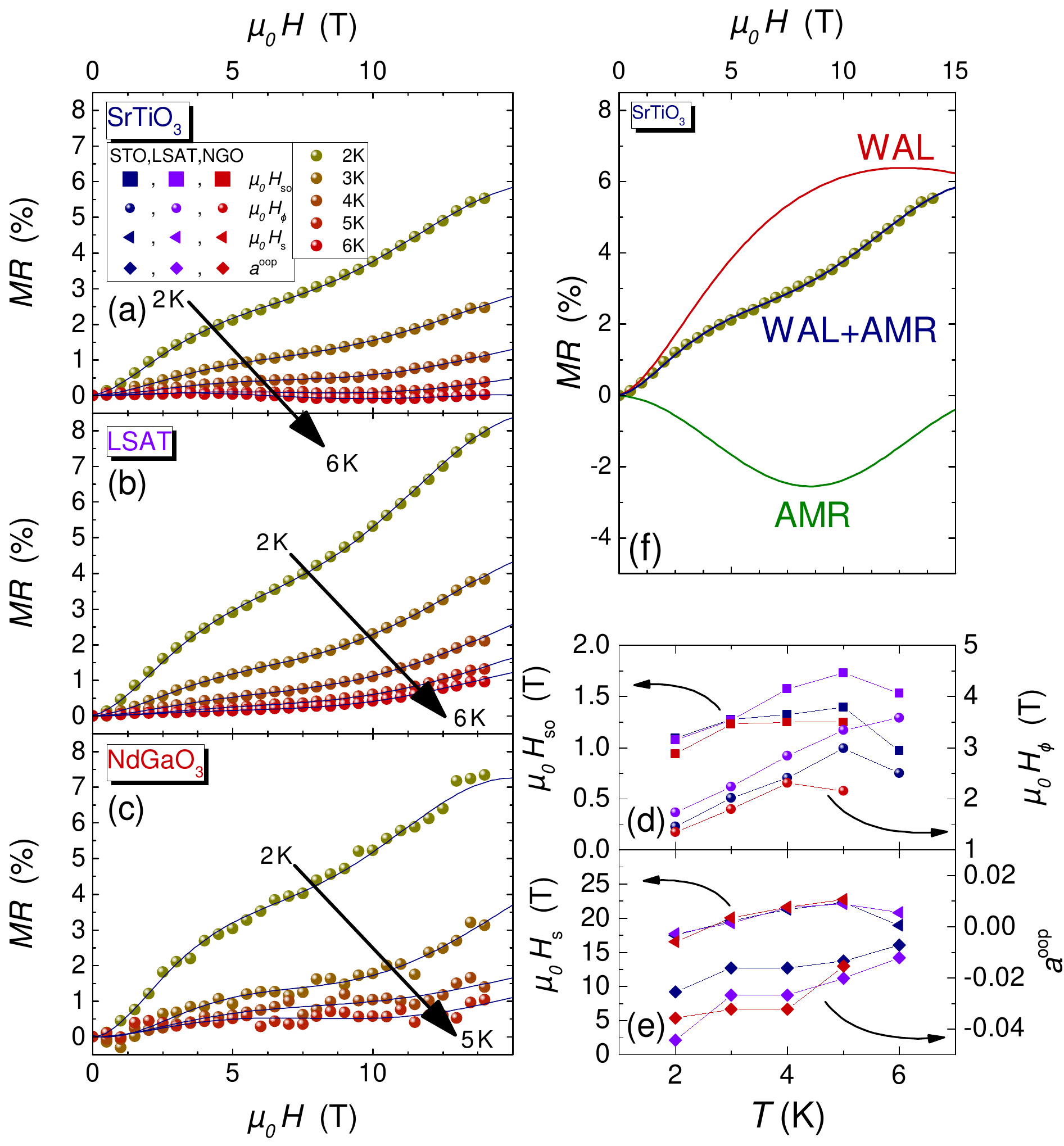}
\caption{\label{Fig_MR-compressive-1}Field dependence of the MR of compressive strained SCIO thin films on STO (a), LSAT (b) and NGO (c) including fits according to equation (\ref{eq:MR:A-typ:1}) -- Scenario I. (d) and (e) Temperature dependence of the fit parameters. (f) Partial contributions of WAL and AMR of the fit of the MR of the sample on STO at $T=2\,\text{K}$.}
\end{figure}

\begin{figure}[t]
\includegraphics[width=\linewidth]{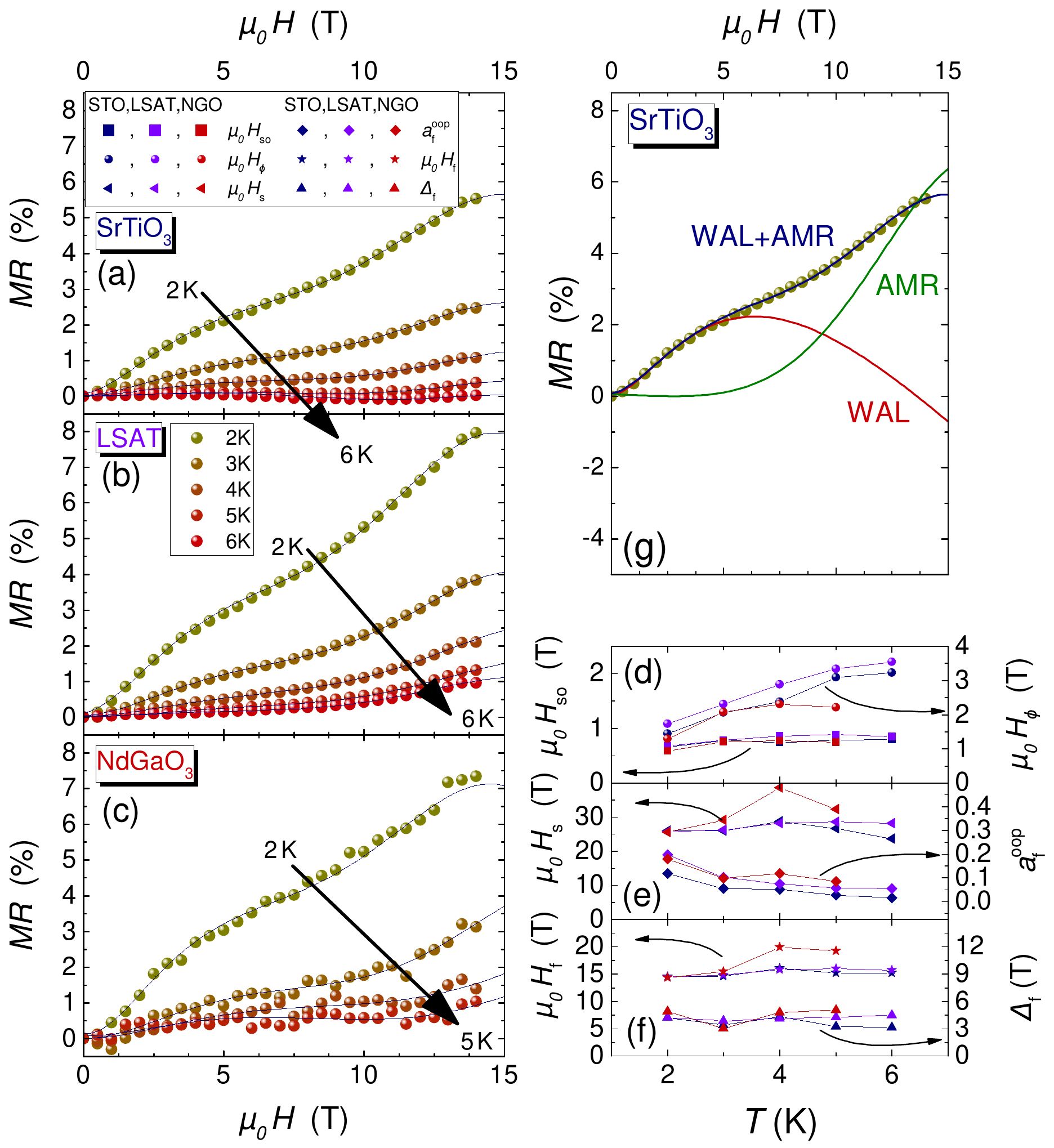}
\caption{\label{Fig_MR-compressive-2}Field dependence of the MR of compressively strained SCIO thin films on STO (a), LSAT (b) and NGO (c) including fits according to equation (\ref{eq:MR:A-typ:2}) -- Scenario II. (d), (e) and (f) Temperature dependence of the fit parameters. $\theta_\text{f} \approx 90\,^\circ$ for all fits. (g) Partial contributions of WAL and AMR of the fit of the MR of the sample on STO at $T=2\,\text{K}$.}
\end{figure}

In the other possible scenario for the (001)$_\text{pc}$ compressive strained SCIO samples we want to use a spin-flip scenario with a spin-flip field $H_\text{f}$ followed by a continues rotation as in the other two cases. In this scenario the magnetisation direction could be modeled with
\begin{widetext}
\begin{align}
	\vec{m}^{(\pm)}_\text{comp. II }(H) =&\vec{e}_\text{oop}\cdot\left[\cos\left(\theta_\text{f}\cdot\frac{|H|-H_\text{s}}{H_\text{s}-H_\text{f}}\right)\cdot\Theta\left(H_\text{s}-|H|\right)\cdot\Theta\left(|H|-H_\text{f}\right)\pm\cdot\Theta\left(H_\text{f}-|H|\right)+\Theta\left(|H|-H_\text{s}\right)\right]\nonumber\\\mp&\vec{e}_\text{ip}\cdot\sin\left(\theta_\text{f}\cdot\frac{|H|-H_\text{s}}{H_\text{s}-H_\text{f}}\right)\cdot\Theta\left(H_\text{s}-|H|\right)\cdot\Theta\left(|H|-H_\text{f}\right)
\end{align}
\end{widetext}
where $\theta_\text{f}$ is the spin-flip angle as defined in Figure \ref{Fig_MR-modell}.
With this the field dependence of the MR in the spin-flip scenarios turns out to be
\begin{align}
	MR \approx &\frac{1}{1+\frac{\rho(0)}{d}\cdot\Delta\sigma(H)}-1\nonumber\\+&a^\text{oop}_\text{f}\cdot\sin^2\left(\theta_f\cdot\frac{|H|-H_\text{s}}{H_\text{s}-H_\text{f}}\right)\cdot\Theta(H_\text{s}-|H|)\cdot\Theta\left(|H|-H_\text{f}\right)\label{eq:MR:A-typ:2:prototyp}
\end{align}
with $a^\text{oop}_\text{f}  = \frac{\mathcal{A}_{11}+\mathcal{A}_{22}-2\mathcal{A}_{33}}{\rho(0)} = -a^\text{ip}$.
Due to fact that we did not observed any sharp jump in the MR data (see Figure \ref{Fig_MR-compressive-1} or \ref{Fig_MR-compressive-2}) the spin-flip field needs to be smeared out by $\Delta_\text{f}$, which is justifiable because due to the incomplete B-site ordering a multi domain structure with different magnetic environments will be develop, resulting in slightly different spin-flip fields.
With a smeared out step function localized at $H = H_\text{f}$ the field dependence of the MR turns out to be

\begin{align}
	MR \approx &\frac{1}{1+\frac{\rho(0)}{d}\cdot\Delta\sigma(H)}-1\nonumber\\+&a^\text{oop}_\text{f}\cdot\sin^2\left(\theta_f\cdot\frac{|H|-H_\text{s}}{H_\text{s}-H_\text{f}}\right)\cdot\frac{\Theta(H_\text{s}-|H|)}{1+\exp\left(-\frac{|H|-H_\text{f}}{\Delta_\text{f}}\right)}\label{eq:MR:A-typ:2}
\end{align}

Both models enable a good description of the collected MR data, as shown in Figure \ref{Fig_MR-compressive-1} and \ref{Fig_MR-compressive-2}.
To select the more realistic scenario we analyze the obtained fitting parameters shown in Figure \ref{Fig_MR-compressive-1} (d) - (e) and \ref{Fig_MR-compressive-2} (d) - (f) in comparison to the values and temperature behaviors obtained for tensile strained samples (see Figure \ref{Fig_MR-tensile} (c) - (d)).

Concerning the WAL part of the MR, the results obtained in the spin-flip scenario (II) are closer to those of the tensile strained case (TSC) than those of the quasi independent FM sublattice scenario (I) and therefore favoring the spin-flip scenario. In more detail the spin-orbit scattering field $H_\text{so}$  in scenario I is twice as big as in TSC and not temperature independent. On the other hand, for the second scenario the magnitude of $H_\text{so}$  is similar to the TS and also its temperature behavior is as expected. The inelastic scattering field $H_\phi$ follows in both cases the same linear behavior as in TSC, but the slope (TSC $\sim 0.36\,\text{T\,K}^{-1}$, I $\sim 0.48\,\text{T\,K}^{-1}$, II $\sim 0.46\,\text{T\,K}^{-1}$) and also the absolute values are in the second scenario closer to those obtained in TSC.

Furthermore the spin-flip scenario is supported by the results of the AMR parts of the MR fits. Certainly strain influences the components $\mathcal{A}_{ij}$ of the galvanomagnetic tensor $\mathcal{A}$ and therefore we did not restrict the prefactors $a^\text{oop}$ and $a^\text{oop}_\text{f}$ for the fits done in Figure \ref{Fig_MR-compressive-1} and \ref{Fig_MR-compressive-2}. However, in principle they are linked to each other by $a^\text{ip} \approx -a^\text{oop}/2 \approx - a^\text{oop}_\text{f}$ and should be therefore positive because $a^\text{ip}$ is negative. In case of the spin-flip scenario $a^\text{oop}_\text{f}$ is positive over the whole temperature range  and at least at $T = 2\,\text{K}$ the relation $a^\text{ip}\approx a^\text{oop}_\text{f}$ is satisfied. Nevertheless the temperature behaviors of $a^\text{ip}$ and $a^\text{oop}_\text{f}$ are slightly different. In contrast to this, the prefactor in scenario I is unexpectedly negative and much smaller (one orde
 r of magnitude) then in scenario II and also the increase of the saturation field $\mu_0H_\text{s}\sim 1.6\,\text{T\,K}^{-1}$ with temperature is in this scenario much stronger than in scenario II ($\sim 0.7\,\text{T\,K}^{-1}$) and in the TSC ($\sim 0.6\,\text{T\,K}^{-1}$).

\begin{figure}[h]
\includegraphics[width=0.9\linewidth]{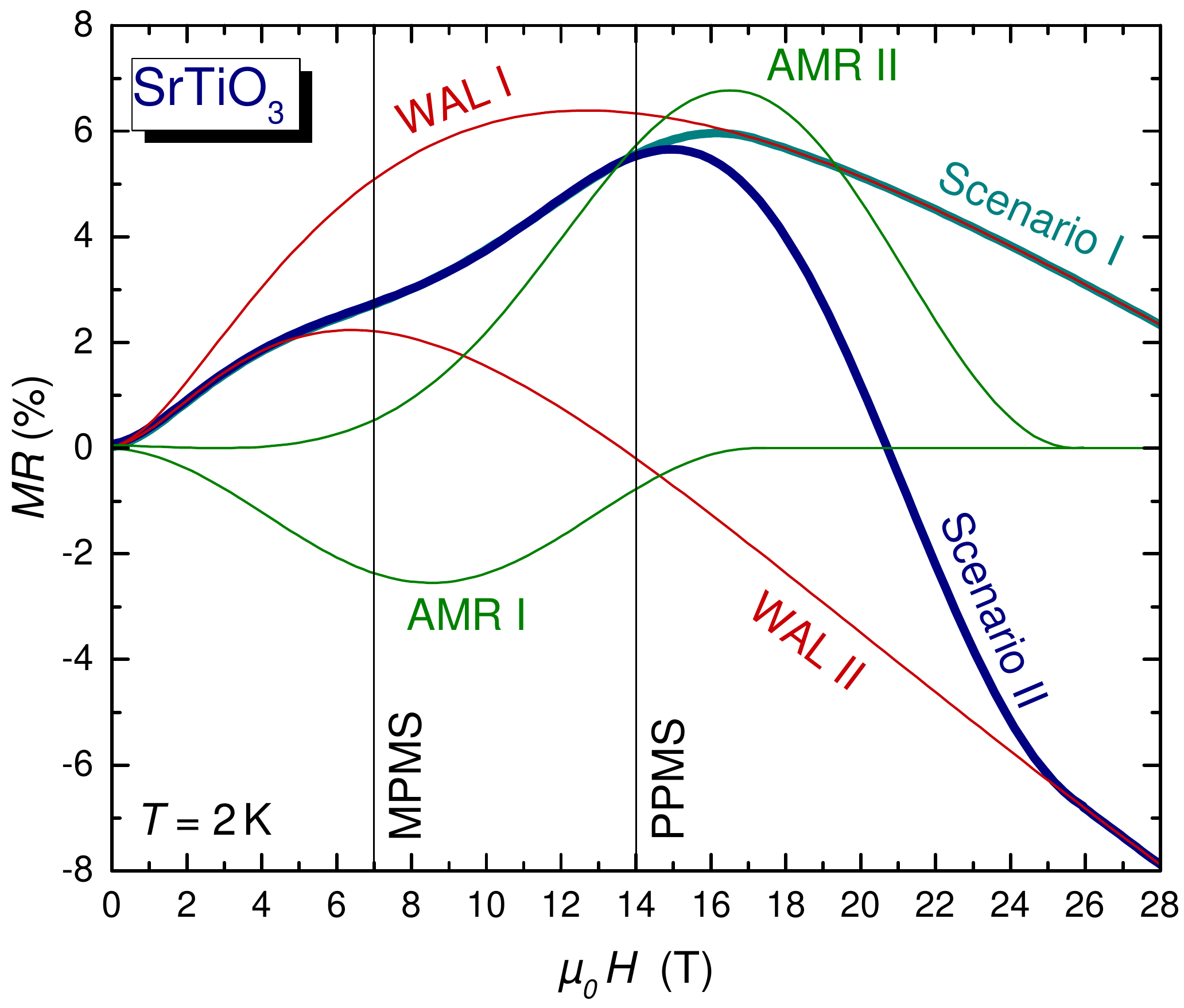}
\caption{\label{Fig_S_MR_STO}(Color online) Simulation of the WAL and AMR part of the total MR signal based on the data of Figure \ref{Fig_MR-compressive-1} and \ref{Fig_MR-compressive-2}.}
\end{figure}

Another argument for the validity of the spin-flip scenario is the temperature independence of the spin-flip field $H_\text{f}$ and the smearing of the spin-flip field $\Delta_\text{f}$.
For a clear separation between the two scenarios we want to propose two experiments:

\begin{enumerate}
	\item By measuring the MR up to fields above the saturation field $H_\text{s}$ the AMR part cancels out of the MR and only the WAL part is remaining (see Figure \ref{Fig_S_MR_STO}). This region could be fitted separatly, using only the WAL formula, equation (\ref{eq:WAL:pure}), and after subtraction of it from the total MR signal the remaining part could be analyzed in both ways.
	\item By using a STO buffer layer on a (001)$_\text{pc}$ piezoelectrical Pb(Mg$_{1/3}$Nb$_{2/3}$)$_{0.72}$Ti$_{0.28}$O$_3$ (PMN-PT) substrate \cite{S_Doerr} it should be possible to build a sample (STO/SCIO/STO/PMN-PT) in which a gate voltage could switch between the two different strain states \textit{in-situ}. This could be used to determine in a first measurement in tensile configuration the WAL part, which than could be subtracted in the second measurement of MR in compressive configuration. As in the first proposed experiment the remaining part could be analyzed in both possible scenarios.
\end{enumerate}

\end{document}